\documentclass[a4paper,12pt]{article}
\pdfoutput=1 

\usepackage{jheppub_X} 

\usepackage{booktabs,colortbl}
\usepackage{cancel,xcolor,multirow}
\usepackage[normal]{caption}
\usepackage{subcaption}
\usepackage{cleveref}
\usepackage{comment}
\usepackage[normalem]{ulem}
\usepackage{tabu}
\usepackage{slashed}
\usepackage{enumitem}
\usepackage{bm}
\usepackage{afterpage}
\usepackage{anyfontsize}

\crefname{table}{Table}{Tables}
\crefname{equation}{Eq.}{Eqs.}
\crefname{appendix}{App.}{Apps.}
\crefname{section}{Sec.}{Secs.}
\crefname{figure}{Fig.}{Figs.}

\usepackage[T1]{fontenc}
\usepackage{listings}
\usepackage[utf8]{inputenc}
\usepackage{lineno}

\makeatletter
\DeclareRobustCommand*{\bfseries}{%
   \not@math@alphabet\bfseries\mathbf
   \fontseries\bfdefault\selectfont
   \boldmath
}
\makeatother

\newcommand{\tev}{~\text{TeV}}
\newcommand{\gev}{~\text{GeV}}

\newcommand{\be}{\begin{equation}} 
\newcommand{\ee}{\end{equation}} 
\newcommand{\bea}{\begin{eqnarray}}  
\newcommand{\eea}{\end{eqnarray}}
\newcommand{\bs}{\begin{split}} 
\newcommand{\es}{\end{split}}

\newcommand{\GeV}{\text{ GeV}}

\newcommand{\TeV}{\text{ TeV}}

\newcommand{\pb}{~\mathrm{pb}}

\newcommand{\MET}{\ensuremath{E_T{\hspace{-0.47cm}/}\hspace{0.35cm}}}

\def\Wboson{\ensuremath{W}}

\def\stopone{\ensuremath{\tilde{t}_1}}

\def\mnino{\ensuremath{m({\footnotesize \ninoone})}}
\def\mstop{\ensuremath{m({\footnotesize \stopone})}}
\def\mt{\ensuremath{m_t}}
\def\mtreco{\ensuremath{m_{t,\rm{reco}}} }

 \def\stop{\ensuremath{\mathchoice%
      {\displaystyle\raise.0ex\hbox{$\displaystyle\tilde t$}}%
         {\textstyle\raise.0ex\hbox{$\textstyle\tilde t$}}%
       {\scriptstyle\raise.0ex\hbox{$\scriptstyle\tilde t$}}%
 {\scriptscriptstyle\raise.0ex\hbox{$\scriptscriptstyle\tilde t$}}}}
\def\stopone{\ensuremath{\mathchoice%
      {\displaystyle\raise.0ex\hbox{$\displaystyle\tilde t_1$}}%
         {\textstyle\raise-.2ex\hbox{$\textstyle\tilde t_1$}}%
       {\scriptstyle\raise.0ex\hbox{$\scriptstyle\tilde t_1$}}%
 {\scriptscriptstyle\raise.0ex\hbox{$\scriptscriptstyle\tilde t_1$}}}}
 
\def\nino{\ensuremath{\mathchoice%
      {\displaystyle\raise.4ex\hbox{$\displaystyle\tilde\chi^0$}}%
         {\textstyle\raise.4ex\hbox{$\textstyle\tilde\chi^0$}}%
       {\scriptstyle\raise.3ex\hbox{$\scriptstyle\tilde\chi^0$}}%
 {\scriptscriptstyle\raise.3ex\hbox{$\scriptscriptstyle\tilde\chi^0$}}}}
\def\ninoone{\ensuremath{\mathchoice%
      {\displaystyle\raise.4ex\hbox{$\displaystyle\tilde\chi^0_1$}}%
         {\textstyle\raise.1ex\hbox{$\textstyle\tilde\chi^0_1$}}%
       {\scriptstyle\raise.3ex\hbox{$\scriptstyle\tilde\chi^0_1$}}%
 {\scriptscriptstyle\raise.3ex\hbox{$\scriptscriptstyle\tilde\chi^0_1$}}}}

\lstdefinestyle{myCustomMatlabStyle}{
  language=Matlab,
  stepnumber=1,
  numbersep=10pt,
  tabsize=4,
  showspaces=false,
  showstringspaces=false
}

\title{\Large 
On the ATLAS Top Mass Measurements and the
Potential for Stealth Stop Contamination \\[5pt]
}

\author[a]{Timothy Cohen,}
\author[a]{Stephanie Majewski,}
\author[a,b]{Bryan Ostdiek,}
\author[a]{and Peter Zheng\hspace{1pt}}
\affiliation[a]{\fontsize{9.3pt}{12pt}\selectfont Institute for Fundamental Science, Department of Physics, University of Oregon, Eugene, OR 97403}
\affiliation[b]{\fontsize{9.3pt}{12pt}\selectfont Department of Physics, Harvard University, Cambridge, MA 02138}

\abstract{
The discovery of the stop --- the Supersymmetric partner of the top quark --- is a key goal of the physics program enabled by the Large Hadron Collider. 
Although much of the accessible parameter space has already been probed, all current searches assume the top mass is known.
This is relevant for the ``stealth stop'' regime, which is characterized by decay kinematics that force the final state top quark off its mass shell; such decays would contaminate the top mass measurements.
We investigate the resulting bias imparted to the template method based ATLAS approach.
A careful recasting of these results shows that effect can be as large as $2.0$ GeV, comparable to the current quoted uncertainty on the top mass.
Thus, a robust exploration of the stealth stop splinter requires the simultaneous consideration of the impact on the top mass. 
Additionally, we explore the robustness of the template technique, and point out a simple strategy for improving the methodology implemented for the semi-leptonic channel.
}

\begin{document}
\maketitle
\flushbottom

\setcounter{page}{2}
\newpage
\section{Introduction}
\label{sec:intro}
%
The top quark plays a critical role in understanding the structure of the Standard Model (SM) and its extensions. The measured value of the top quark mass $m_t$ (and Yukawa coupling) is an important input for precision tests of the self consistency of the SM.
If nature is Supersymmetric (SUSY), the top should have a partner --- the stop --- that tames the ultraviolet sensitivity implied by the coupling between the top quark and the Higgs boson.  
Since this is one of the most compelling ways to extend the SM, an extensive search program for the stop has been conducted by both ATLAS~\cite{Aaboud:2017ayj,Aaboud:2017aeu, Aad:2020sgw} and CMS~\cite{Sirunyan:2017wif,Sirunyan:2017xse}, yielding an impressive exclusion covering stop masses as high as $\sim1.2\,\text{TeV}$.
Although the narrow ``splinter'' region ($\mstop \sim m_t$) now appears to be closed, all current searches assume that $m_t$ is known. As we will argue here, this assumption deserves further scrutiny.

The SUSY framework makes it manifest that as the mass of the stop becomes parametrically large with respect to the weak scale, the fundamental parameters become increasingly fine tuned in order to reproduce the measured Higgs vacuum expectation value. 
Thus, there remains significant interest in this inherently natural but notoriously difficult to explore ``stealth'' stop region of parameter space. 
The degeneracy of these mass parameters implies tight kinematic constraints such that the final state looks nearly identical to top pair production, albeit where the tops are off-shell. 
Thus, not only does the presence of copious SM top pair production obscure the presence of the stop, but if the stop exists with a mass in this regime, the precision measurements of the top mass itself would be biased due to the presence of stop decays.
We build upon previous studies of the subtle phenomenological signals of stealth stops~\cite{Czakon:2014fka,Eifert:2014kea} in a number of ways: we study the impact on all three channels (all-hadronic, semi-leptonic, and di-leptonic) tracking their correlated effects, we use the most up-to-date ATLAS measurements, and we recast the template method in detail and study its robustness.
Our main results quantify the potential contamination of these measurements due to a stealth stop. 
Furthermore, we propose an improvement in the methodology for measuring the top mass by explicitly using both tops in the event; we highlight this in the semi-leptonic channel.

\begin{figure}[t]
\center
\includegraphics[width=\linewidth]{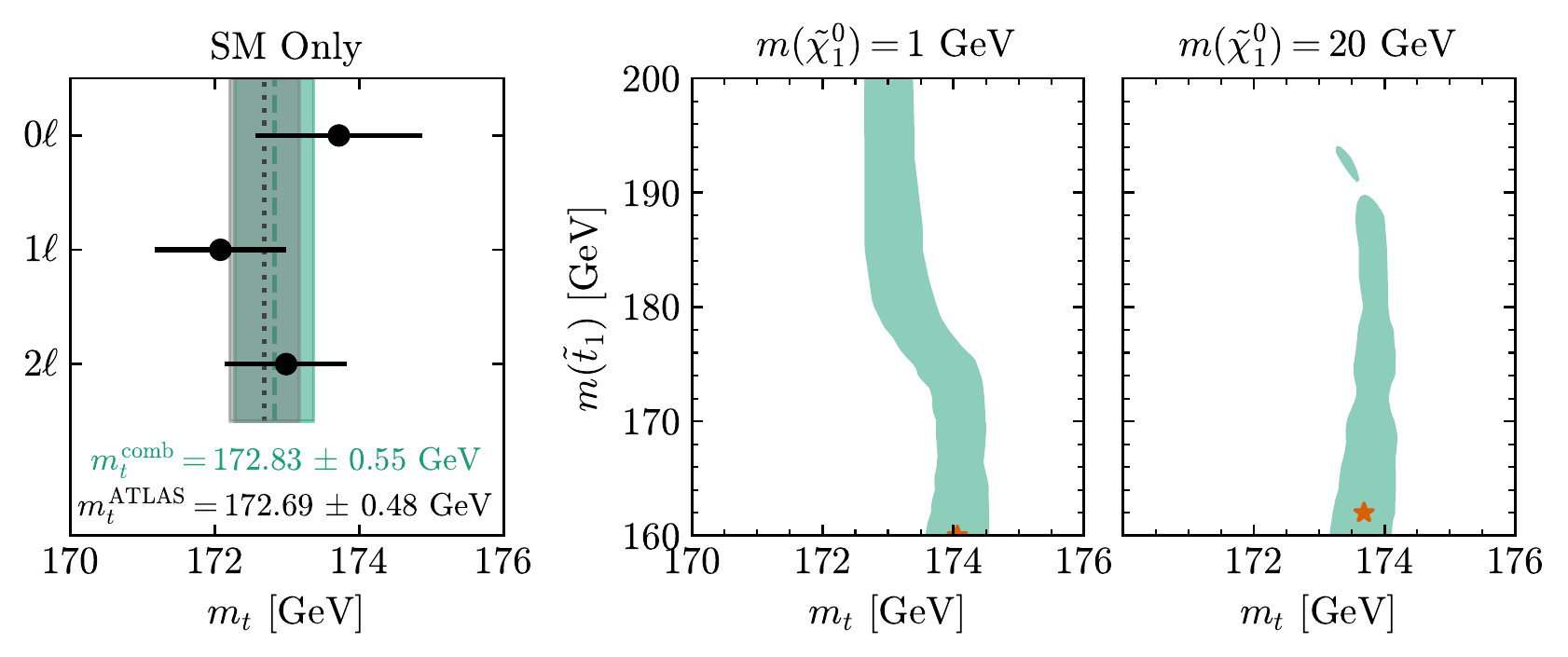}
\caption{
Parameters which give the best fit for the top mass measurement by ATLAS by combining the three different channels at $\sqrt{s}=8\tev$. The ATLAS measurements of the three different channels are shown in the left panel: all-hadronic~\cite{Aaboud:2017mae} $(\mt = 173.72 \pm 1.15 \gev)$, semi-leptonic~\cite{Aaboud:2018zbu} $(\mt = 172.08 \pm 0.91 \gev)$, and di-leptonic~\cite{Aaboud:2016igd} $(\mt = 172.99 \pm 0.84 \gev)$.
Assuming the SM only, the green band shows our crude best-fit value of $\mt^{\mathrm{comb}}$ with its associated uncertainty and the black band shows the ATLAS combination $\big(\mt^{\mathrm{ATLAS}}\big)$ given in Ref.~\cite{Aaboud:2018zbu}, taking into account 7 and 8 TeV data.
The center and right panels illustrate the impact of stealth stop contamination, where we show the best fit point in the $\mt$ - $\mstop$ plane (marked by the orange star), and the 1-$\sigma$ confidence interval as shaded bands.
When the stops must decay through an off-shell top, they shift the reconstructed template mass extraction to values that are smaller than the actual top mass chosen in the Monte Carlo event generation.}
\label{fig:SMCombination}
\end{figure}

To achieve this goal, we carefully recast precise measurements of the top quark mass, choosing the ATLAS Collaboration's template method for its straightforward response to stop signal contamination; we expect our results to be generally applicable regardless of the method used.\footnote{The top mass has also been precisely measured by the CDF~\cite{Abazov:2015spa} and D0~\cite{D0:2016ull} Collaborations using the matrix element technique, and by the CMS Collaboration using the ideogram method~\cite{Khachatryan:2015hba,Sirunyan:2017idq}.}
The ATLAS Collaboration characterizes its measurements by the top decay products considered: all-hadronic~\cite{Aaboud:2017mae} (with no leptons in the final state), semi-leptonic~\cite{Aaboud:2018zbu} (where one top quark decays to jets and the other decays via an electron or muon), and di-leptonic~\cite{Aaboud:2016igd} (where both top quarks decay via an electron or muon). These measurements are based on the $\sqrt{s}=8\TeV$ dataset and are summarized in the left panel of \cref{fig:SMCombination}; we also provide a crude combination\footnote{The details of our naive approach for combining measurements are presented in \cref{Sec:Combined}.} assuming uncorrelated Gaussian error bars, which is consistent with the sophisticated combination performed by ATLAS that includes the $\sqrt{s}=7\tev$ dataset. A general review of the template method together with an illustrative toy example is presented in \cref{sec:tpl}. In \cref{sec:semi}, we present a detailed analysis of the semi-leptonic channel and propose an improved strategy that requires minor modifications to the current ATLAS approach. 

\begin{figure}[t]
\begin{center}
\includegraphics[width=0.45\linewidth]{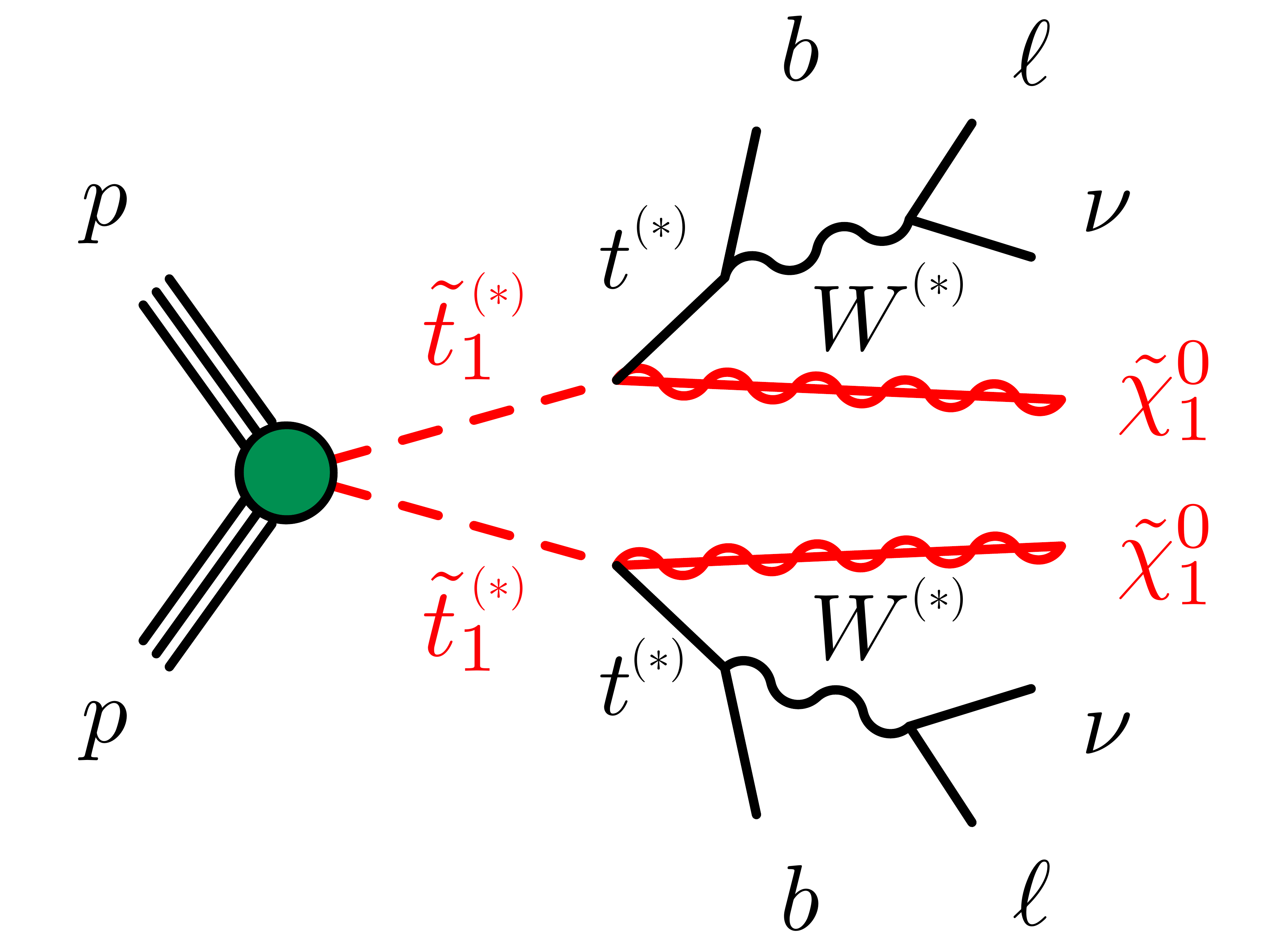}%
\hspace{30pt}
\includegraphics[width=0.45\linewidth]{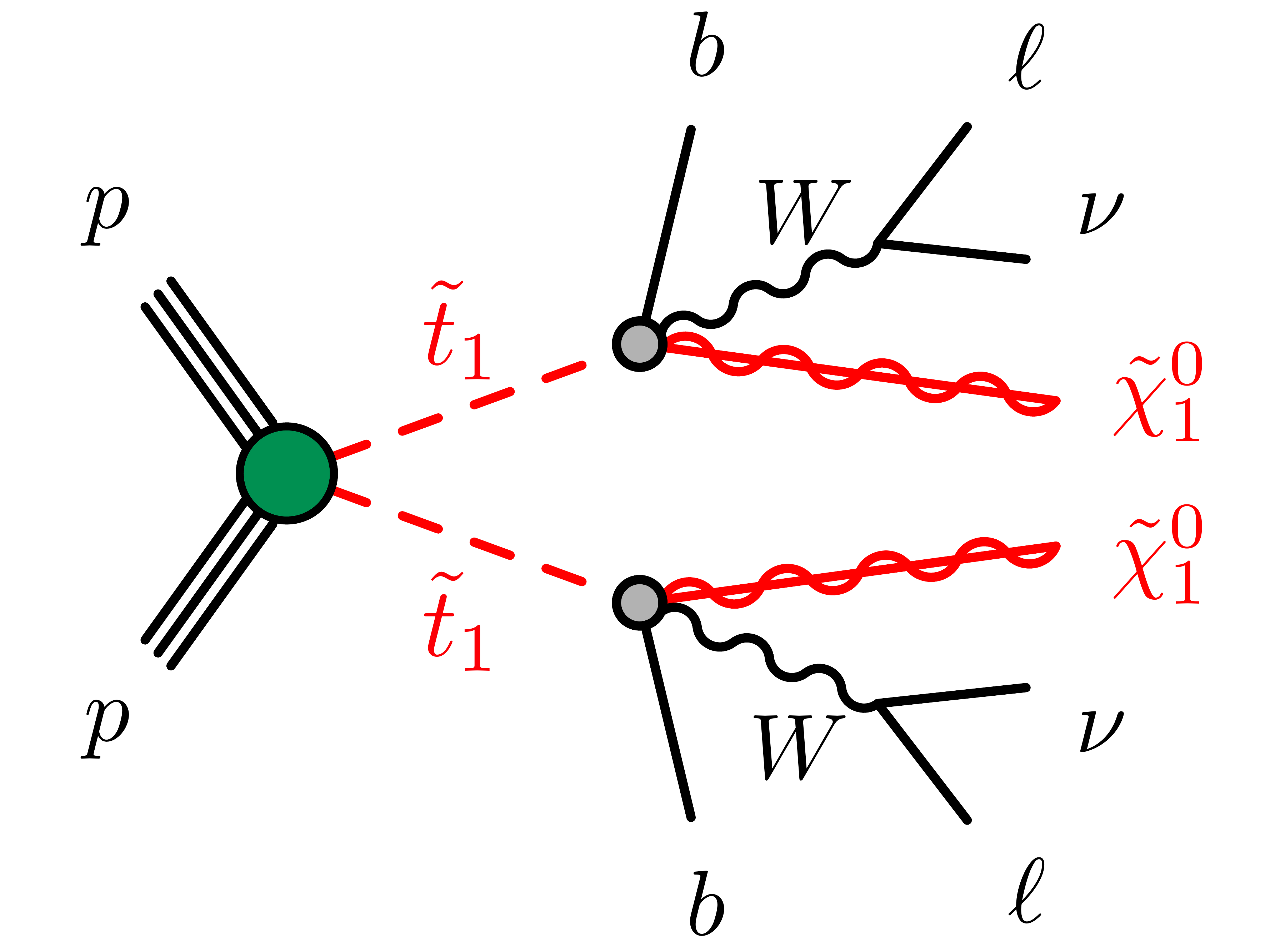}%
\caption{The left diagram illustrates the full process for the parameter space where $\mstop -\mnino < m_t$ including the off-shell propagators which encode the non-trivial kinematic correlations, and the right diagram illustrates the same process in the narrow width approximation.  The green circles represent the full tree-level stop pair production matrix element, which is included in our simulations.  The gray circles represent decays that do not include any matrix element information, \emph{i.e.}, the particles are decayed using phase space alone. The superscripts $(*)$ denote particles that can go off shell.  This figure was adapted from diagrams given in Ref.~\cite{Aaboud:2017nfd}.}
\label{fig:FeynmanDiagrams}
\end{center}
\end{figure}

The potential contamination from a stealth stop is modeled using the ``stop-neutralino'' Simplified Model~\cite{Papucci:2011wy, Essig:2011qg, Alves:2011wf}, which is inspired by the ``more minimal SUSY SM''~\cite{Dimopoulos:1995mi, Cohen:1996vb}. Under the well-motivated assumption the lightest superpartner is a stable state, phenomenological viability requires that the particle be neutral, thereby providing a dark matter candidate~\cite{Fayet:1976et, Jungman:1995df}, the so-called lightest neutralino $\ninoone$. The rate of direct stop pair production is fully specified by $\mstop$, and each stop subsequently decays to an on- or off-shell top quark and $\ninoone$, as illustrated in \cref{fig:FeynmanDiagrams}.\footnote{When the stop decays involve off-shell tops, the final state branching ratio for the stop can recieve non-trivial contributions from other channels as discussed in Ref.~\cite{Krizka:2012ah}.  Since this depends on the details of the underlying SUSY breaking flavor structure, we will ignore these subtleties and assume that the stop decays to an off-shell top and a neutralino 100\% of the time.} The stealth stop region of parameter space is thus more precisely defined by $\mstop - m_t \simeq \mnino$. The degeneracy of these mass parameters implies tight kinematic constraints such that the final state looks nearly identical to top pair production, perhaps with some additional missing energy due to the presence of the neutralino. There have been many phenomenological studies to constrain light or compressed stops, \emph{e.g.}~\cite{Essig:2011qg, Papucci:2011wy, Alves:2012ft, Han:2012fw, Kilic:2012kw, Krizka:2012ah,
Agrawal:2013kha, Dutta:2013gga,
Aebischer:2014lfa, Cho:2014yma, Buckley:2014fqa, Grober:2014aha, Fan:2014txa,
An:2015uwa, Batell:2015koa, Batell:2015zla, Belyaev:2015gna, Fan:2015mxp, Ferretti:2015dea, Hikasa:2015lma, Macaluso:2015wja,
An:2016nlb, Cheng:2016npb, Konar:2016ata,
Aebischer:2017aqa,
Cohen:2018arg, Roxlo:2018adx}.
The stop is off-shell in much of this parameter space; a careful modeling of the angular distributions of the final state is needed since the kinematics can have a non-trivial impact on the resulting efficiencies. 
Therefore, one must abandon the narrow-width approximation~\cite{Gigg:2008yc} (depicted in the right panel of \cref{fig:FeynmanDiagrams}) and compute the full four-body kinematics (as illustrated in the left panel of \cref{fig:FeynmanDiagrams}).
Here, we will follow the procedure developed in Ref.~\cite{Cohen:2018arg} for simulating events including these effects.

The central and right panels of \cref{fig:SMCombination} provide a summary of our main results, which are described in detail in \cref{sec:result}. We introduce stop contamination into the recasted top mass measurements and provide a simple combination of the three channels for the 1-$\sigma$ best-fit region in the $\mstop$ - $m_t$ plane. The best fit point is shown as an orange star. Two neutralino mass points are shown: $\mnino = 1 \text{ GeV}$ (center) and $\mnino = 20 \text{ GeV}$ (right) for a range of stop and top masses. The maximum bias for each channel is also summarized in \cref{table:maxbias}. The bias in the observed top mass depends on the mass of the contaminating stops, and it can be as large as $2.0\GeV$ in the di-leptonic channel.

\begin{table}[t]
\centering 

\renewcommand{\arraystretch}{1.3}
\setlength{\tabcolsep}{20 pt}
\begin{tabular}{c c c c} 
\toprule
 & All-hadronic & Di-leptonic & Semi-leptonic\\ 
\hline 
\mstop&  172.2 GeV& 166.5 GeV&160.8 GeV\\ 
Bias & $-0.5$ GeV & $-2.0$ GeV & $-1.3$ GeV \\
\bottomrule
\end{tabular}
\caption{Summary of the maximum bias on the measured $m_t$ due to stop contamination in each channel, assuming $\mnino = 1 \text{ GeV}$. The top row shows the mass of the stop that maximally biases the experimentally measured mass from the Monte Carlo truth mass. The size of the bias in the measurement for each channel is shown in the bottom row. 
}
\label{table:maxbias} 
\end{table}

Our results have an important impact on interpretations of stop exclusion in the stealth stop region. 
Both precision measurements and direct searches have attempted to whittle away the apparent available parameter space to a mere ``splinter.''  
An early ATLAS approach to examining this region relied on precision measurements of the top cross section~\cite{Aad:2014kva}, although only results for $\mnino = 1 \text{ GeV}$ were presented. 
This motivated our previous study~\cite{Cohen:2018arg}, where we performed a careful recasting of the ATLAS exclusion to extend it into the full stop-neutralino mass plane. 
More recently, both ATLAS~\cite{Aaboud:2019hwz} and CMS~\cite{Sirunyan:2019zyu} have exploited the clean signature and angular distributions of $e\mu$ events, nearly excluding the narrow splinter-like region. 
However, Ref.~\cite{Aad:2015pfx} shows that observed limits on the stop mass using the $t\bar{t}$ cross section ratio at 7 and 8 TeV center-of-mass energy collisions at the LHC drop from around 180 GeV to 160 GeV if the top mass is changed from 172.5 to 175 GeV, indicating that $\mathcal{O}(1\GeV)$ shifts in the top mass can have an appreciable impact on the stop limits. In this paper we demonstrate that stealth stops can contaminate the top mass measurement at this level, which would lead one to infer that the top mass is lighter than its true underlying value. 
To know if we have actually closed the window on light stops, the interplay between the measured top mass and the stealth stop exclusion limits must be rigorously explored.

\section{The Template Method}
\label{sec:tpl}
Any discussion relating the theoretical mass of a particle to an experimental observable requires care. 
From a quantum field theory point of view, the choice of scheme is defined by how one decides to remove the UV divergences when renormalizing perturbation theory.
Two common choices yield what is referred to as the ``pole'' mass or the ``$\overline{\text{MS}}$'' mass.
In the three measurements studied here, ATLAS avoids these issues and instead infers what is often called the ``Monte Carlo'' mass by comparing some observable that is sensitive to the top mass against Monte Carlo generator predictions as a function of the numerically implemented mass parameter.
The MC top quark mass $m_{t,\rm{MC}}$ is related to the field-theoretic pole mass $m_{t,\rm{pole}}$ as
\begin{equation}
m_{t,\rm{MC}} = m_{t,\rm{pole}} \pm \delta m_t\,.
\end{equation}
In the discrepancy $\delta m_t\sim\mathcal{O}\big(Q_0 ~\alpha_s (Q_0)\big)$, $Q_0$ corresponds to the scale of the shower cutoff \cite{Fleming:2007qr, Hoang:2008xm, Butenschoen:2016lpz, Hoang:2017kmk, Hoang:2018zrp}, and $\alpha_s$ is the strong coupling. 
Other studies suggest the uncertainty in this conversion is on the order of the hadronization scale \cite{Nason:2016tiy, Nason:2017cxd}; see Ref.~\cite{Andreassen:2017ugs} for a study on reducing this ambiguity by means of jet grooming.
We conclude that the difference is generally on the order of a few hundred MeV, which is comparable to typical modern experimental precision~\cite{Corcella:2019tgt}. 
From here forward, we will put these issues aside and focus on the methodology employed by ATLAS --- we emphasize that $\delta m_t$ is another source of systematic uncertainty that must be tracked when comparing the value of $m_t$ measured by ATLAS measurement to other approaches or as an input to a theory calculation.

In order to compare Monte Carlo predictions to data, ATLAS relies on a template method.
An observable $O$ is chosen such that it is sensitive to the top mass, and simulations are then used to compute distributions for multiple values of $m_t$.
Clearly, the particular choice of $O$ depends on the channel under consideration; for example, in the all-hadronic channel~\cite{Aaboud:2017mae}, ATLAS constructs the ratio between the 3- and 2-jet invariant masses as this minimizes sensitivity to the jet energy scale uncertainty.
Samples of the distributions for $O$ are generated over a range of values for the top mass (ATLAS does this for five $m_t$ values from 167.5 GeV to 177.5 GeV).
A set of preselection cuts are then applied, and each resulting distribution is fit with the same parametric curve.
Then the resulting best-fit values are assumed to be linear functions of $m_t$, and an interpolation as a function of $m_t$ is derived by linearly fitting the parameter variations as a function of $m_t$.
This resulting object is the so-called template, which allows one to ``predict'' the shape of $O$ as a function of $m_t$.
To make this procedure more concrete, and to highlight some of its features, we work out a detailed toy example template in what follows.

\subsection{A Toy Example}
\label{sec:toy}
In this section, we present a toy model that illustrates how the template method works in practice. 
For now, we will assume that the distribution for the observable $O$ has a characteristic peak followed by an extended tail. 
For concreteness, we model such a shape using a Gaussian for the peak and a Landau function for the tail, where the latter is defined as
\begin{equation}
P_{\rm{\,Landau}}\big(x;\mu, c \big) = \frac{1}{\pi\, c}\int_0^\infty  \text{d}t ~ e^{-t}\cos\Bigg[t\left(\frac{x-\mu}{c}\right) + \frac{2\, t}{\pi}\log\left(\frac{t}{c}\right)\Bigg]\,,
\label{eqn:Landau}
\end{equation}
where $\mu$ essentially controls the location of the peak and $c$ controls the width of the distribution.  
This toy model is described by six parameters: 
\begin{equation}
P\big(x; a, b, \chi, \sigma, \mu, c\big) = a\, P_{\,\rm{Gaussian}}\big(x; \chi, \sigma\big) + b\,P_{\,\rm{Landau}}\big(x;\mu, c\big)\,.
\label{eqn:toy_full}
\end{equation}
The parameters $a$ and $b$ control the relative normalizations of the Gaussian and the Landau components,\footnote{As a probability distribution, the values of $a$ and $b$ should be chosen such that the distribution integrates to unity.  We do not enforce this constraint since we do not sample over the full allowed range of values for $x$, and furthermore, we found this more flexible form yields better fits.} while $\chi$ and $\sigma$ are the mean and standard deviation of the Gaussian, respectively.   
We choose to use a Gaussian plus a Landau as our toy distribution since this is the shape used by ATLAS for the di-leptonic measurement. The other two channels are fit to similar distributions as discussed below.

The key to choosing a good observable is that its shape (and ideally the location of a peak) must change as a function of the underlying parameter of interest --- for the measurements of interest below, this parameter is the top mass, while in the toy model studied in this section, we will call this $m_\text{toy}$.
We model the ``truth-level'' change in the underlying six parameters defined in \cref{eqn:toy_full} as linear functions of $m_\text{toy}$, which are chosen to closely mimic those that ATLAS extracts from real data.

\begin{figure}[t]
\begin{center}
\includegraphics[width=0.85\linewidth]{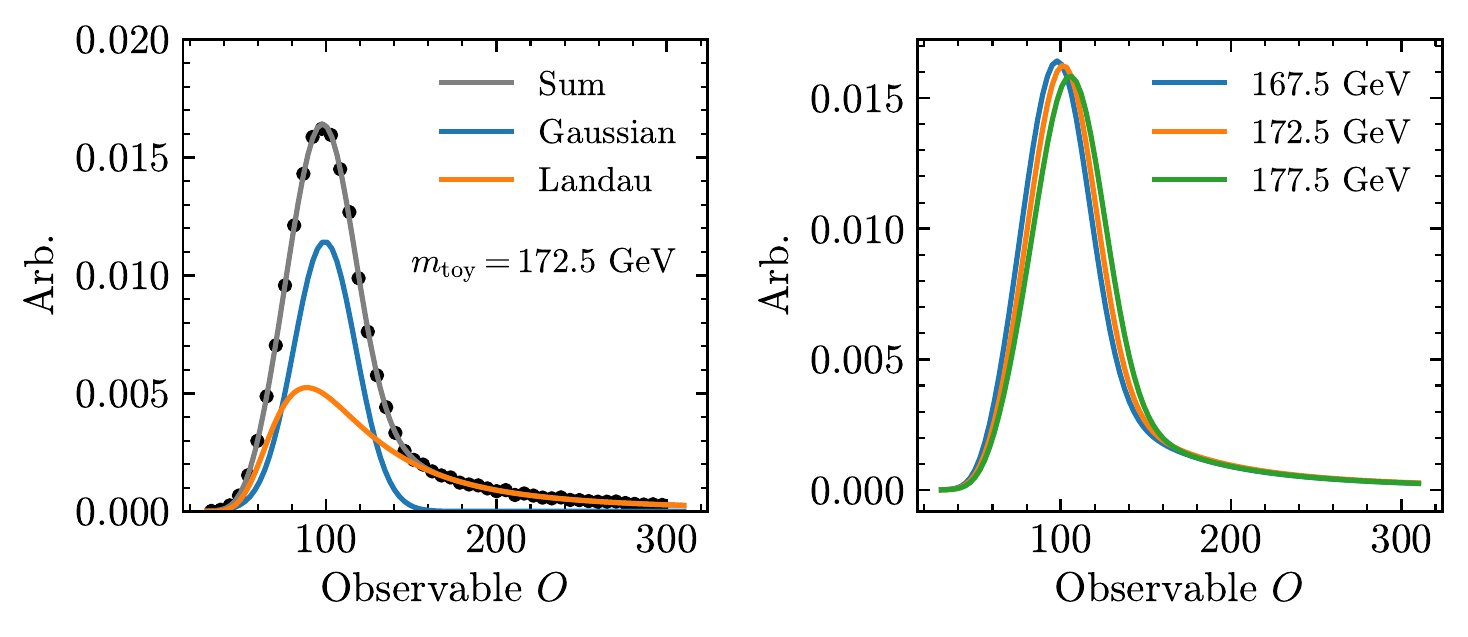}%
\caption{
Our toy observable is modeled with a probability distribution which is the sum of a Gaussian for the peak and a Landau function for the tail.
The gray curve in the left panel shows the fit to the black data points, while the blue and orange lines show the individual contributions of the Gaussian and Landau components, respectively. 
The right panel demonstrates how the shape of the observable changes as a function of $m_\text{toy}$: the location of the peak of the observable $O$ is highly correlated with $m_\text{toy}$. 
} 
\label{fig:toyplot1}
\end{center}
\end{figure}

Once the observable $O$ and the parametric model are chosen, the next step is to construct the templates. 
The ATLAS approach relies on Monte Carlo simulations for different choices of $m_t$. 
For our toy example, we draw samples from the truth-level probability distributions at five values of $m_\text{toy}$ using the Metropolis-Hastings Markov Chain Monte Carlo (MCMC) algorithm~\cite{Metropolis, Hastings}. 
A dataset of 10,000 elements is constructed for each choice of $m_\text{toy}$, which are subsequently binned and normalized.  
We then fit the resulting histograms to the distribution given in \cref{eqn:toy_full}.  
An example fit is shown in the left panel of \cref{fig:toyplot1}, comparing the fitted distribution to the toy data, and the right panel displays the best-fit templates for three different values of $m_\text{toy}$.

\begin{figure}[t]
\begin{center}
\includegraphics[width=0.85\linewidth]{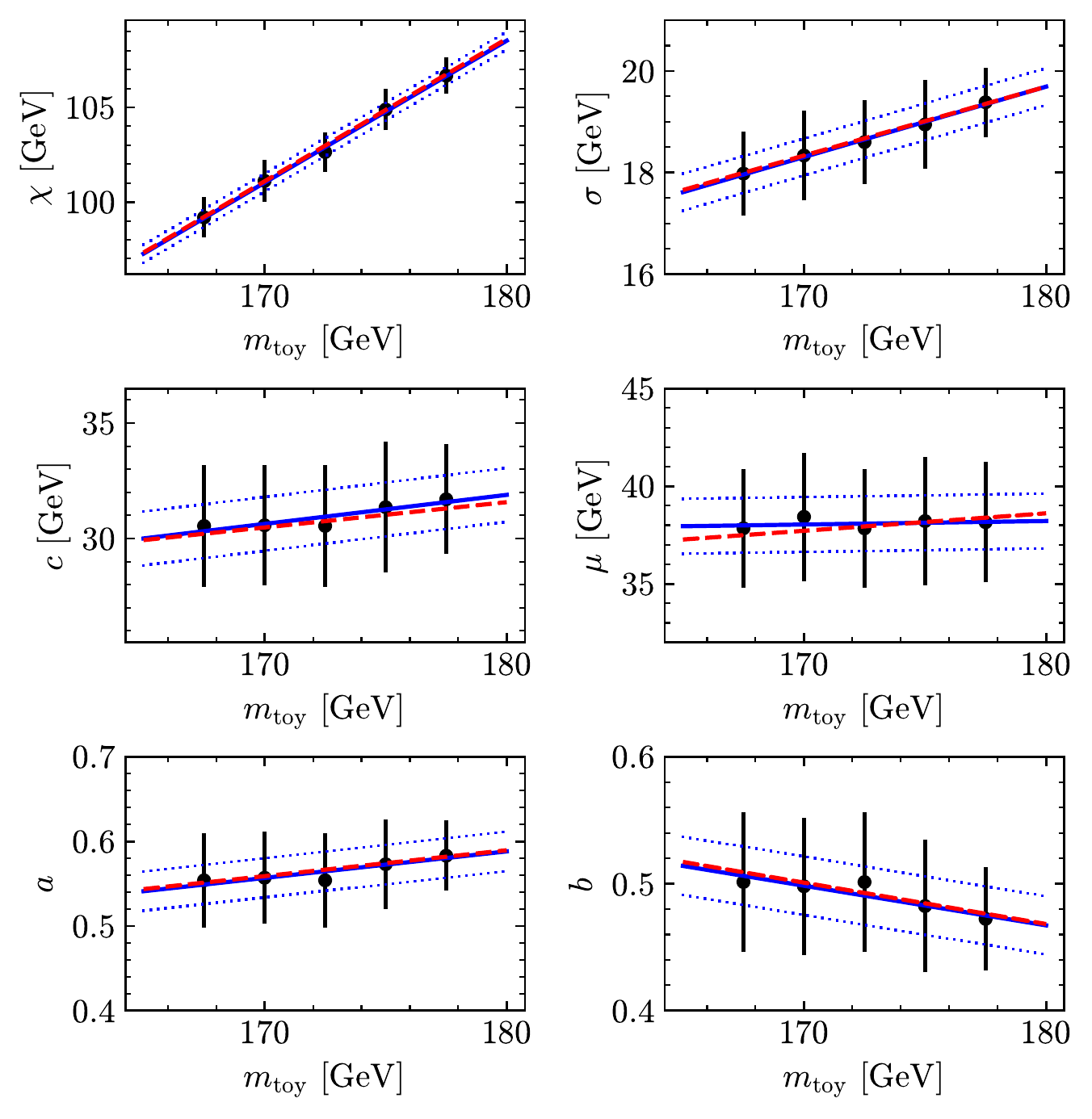}%
\caption{
The toy observable is fit to the six-parameter function in \cref{eqn:toy_full}. 
We take the dependence of each parameter as a function of $m_\text{toy}$ from the red dashed lines. 
For each value of $m_\text{toy}$, a dataset is drawn from these \emph{truth}-level probability distributions, which is subsequently fit to the function in \cref{eqn:toy_full}.
This process is repeated 100 times; the black dots with error bars provide the mean and standard deviation inferred for each parameter. 
The blue solid lines give the resultant linear fits to these points with associated errors.
The upper and lower dotted lines in each panel are the result of systematically shifting the best fit line up and down by the covariance of the $y$-intercept as determined during the fitting procedure (see \cref{sec:toy}). 
The envelope of dotted lines provides some insight into the systematic uncertainty associated with poorly modeled templates. 
}
\label{fig:toyplot2}
\end{center}
\end{figure}

In order to account for the statistical noise due to finite sample sizes, we generate 100 independent data sets from the truth-level distribution, and find the best fit parameters for each.
The mean and the standard deviation for each of the parameters are shown as the data points with error bars in~\cref{fig:toyplot2}, and the linear functions are indicated by the dashed red lines.  
It is not surprising to see that the largest range occurs for the variable $\chi$, since this determines location of the peak. 
Additionally, this parameter $\chi$ has the smallest fractional uncertainty of $\sim 1\%$, while the other parameter error bars vary from $\sim 4\%$ to as much as $\sim 8 \%$, which can be traced back to its sensitivity to the position of the peak of the distribution. 
Each of these distributions is then fit to a line including the impact of the error bars on the fit, as shown by the blue lines in~\cref{fig:toyplot2}.

The final step for constructing a template is to use these linear fits as a function of $m_\text{toy}$ to convert the parametric model of \cref{eqn:toy_full} into a function of $m_\text{toy}$ alone. 
Explicitly, the model becomes $P\big(x; a, b, \chi, \sigma, \mu, c\big) \rightarrow P\big(x; m_\text{toy}\big)$, where each of the original parameters is determined by the appropriate best fit linear function of $m_\text{toy}$.  
Finally, one can use these templates to extract a mass measurement by fitting the template (which is now a function of the single parameter $m_\text{toy}$) to the experimentally determined distribution.

We identify two sources of uncertainty within the template method as implemented here: the first is the statistical uncertainty from using the derived template, and the second is the systematics associated with deriving the template itself.
We use a closure test to assess the size of these uncertainties.
An extra 100 sets of samples for a given mass point are generated; for each we find a template mass that best fits the distribution (using the templates of the blue lines of \cref{fig:toyplot2}).
The difference between the truth and extracted values are small, and the standard deviation gives us an estimate for the uncertainty of using the template, around 0.33 GeV for the toy model.

To measure the second source of uncertainty, which results from the assumption of the linear dependence of the template parameters, we repeat the closure test using the dotted lines of \cref{fig:toyplot2} denoting the uncertainty of the linear fits.
Using the shifted template results in extracting $m_{\text{toy}} \sim m_{\text{true}} \pm 1.5\gev$, depending on if the upper or lower shift is considered.
In the toy model, we find this systematic uncertainty from deriving the template is larger than the statistical uncertainty.
When we perform the same tests for the ATLAS top mass measurements, we find that the two uncertainties are similar in size to each other and subdominant to other quoted experimental uncertainties, \emph{e.g.} that come from the parton distribution functions or the jet energy scale.

\subsection{Dependence on the Choice of Fit Function}
\label{sec:choice}
The last section addressed some of the uncertainty associated with constructing a template. 
However, in performing those tests we used a parametric fit function that has the exact same form as the true underlying distribution. 
This is in contrast with the fit functions utilized by ATLAS, which are not necessarily determined from the underlying physics. 
As we will show here, the template method is quite robust as long as the model parameters are linearly dependent on $m_\text{toy}$, even if the model does not provide a particularly good fit to the distribution of the observable $O$.

\begin{figure}[h!]
\begin{center}
\includegraphics[width=0.85\linewidth]{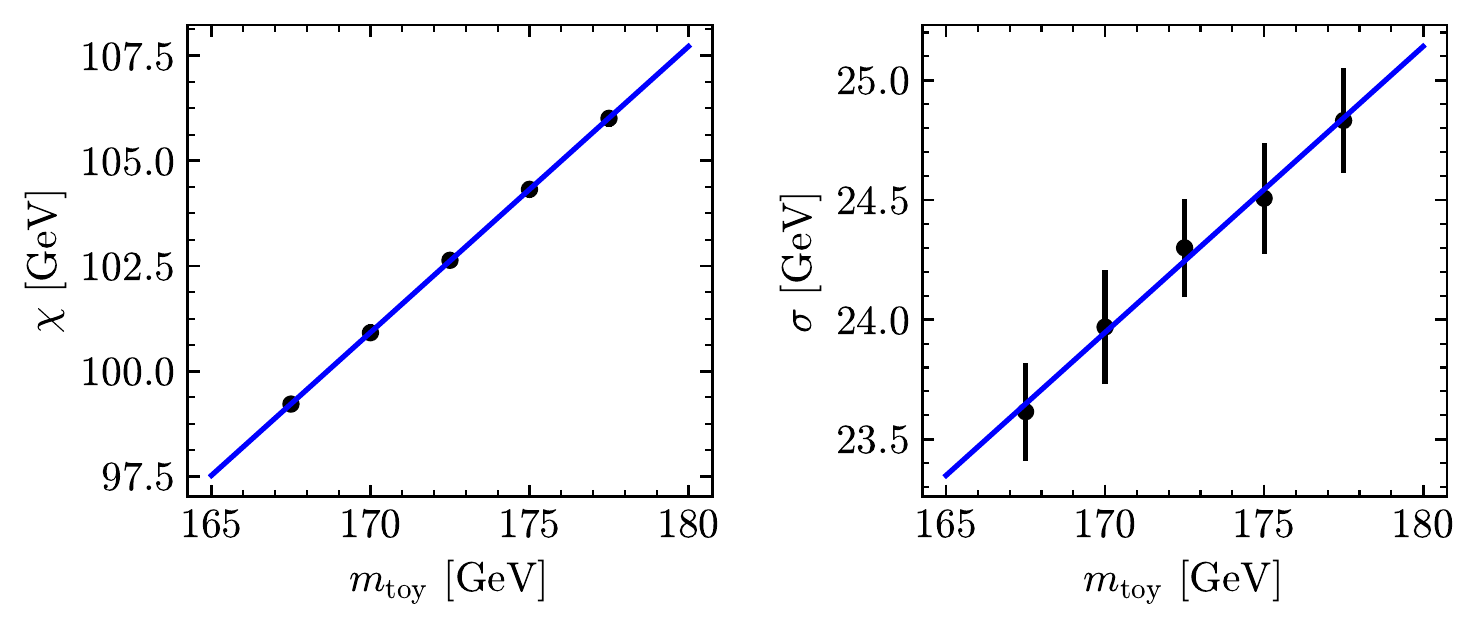}%
\caption{
The results of a test using the toy model, where the parametric fitting function was assumed to be a Gaussian alone, which does not provide a good fit to the underlying $O$ distribution.
The left and right panels show the mean and standard deviations of the template Gaussian for different values of $m_\text{toy}$ along with the linear fit used to generate the template.
The data points and error bars come from the average and standard deviation of 100 fittings of random samples.
}
\label{fig:closeGaus}
\end{center}
\end{figure}

\begin{figure}[h!]
\begin{center}
\includegraphics[width=0.85\linewidth]{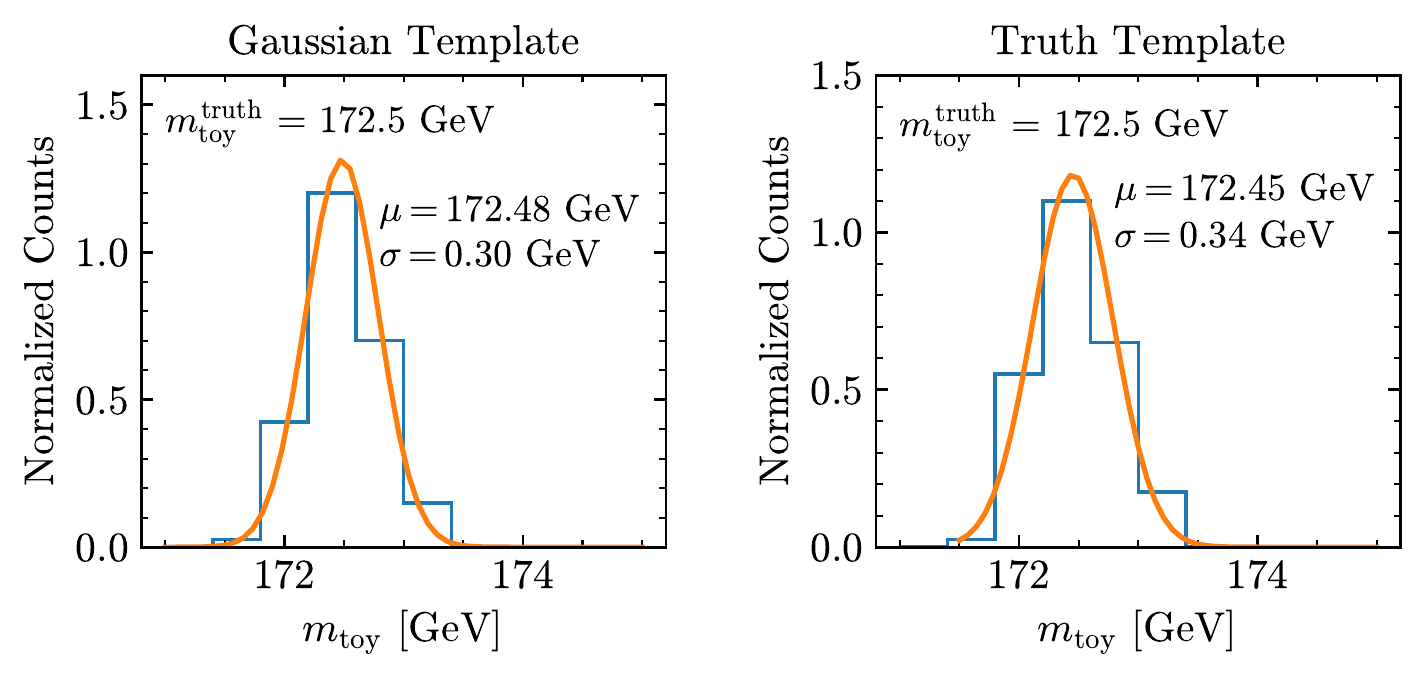}%
\caption{
The left panel shows the value extracted from 100 independent data sets of $m_\text{toy}=$ 172.5 GeV assuming this Gaussian based template while the right panel shows the same data with the template from \cref{eqn:toy_full}.
The bias and uncertainty with the simpler model are smaller than when we used the \emph{truth} model. 
}
\label{fig:closeGaus2}
\end{center}
\end{figure}

To illustrate this point, we repeat the template analysis with the true data distributed according to the same toy model described by \cref{fig:toyplot2}. 
We now fit the distributions by a Gaussian alone, which does not model the tail of the $O$ distribution.
As shown in \cref{fig:closeGaus}, the best-fit Gaussian tracks the location of the peak, which is highly correlated with the underlying $m_\text{toy}$.
We then generate a template in analogy with above, and perform a closure test, which yields the left panel of~\cref{fig:closeGaus2}. As comparison, we provide the closure test result from the truth template on the right pannel. 
Surprisingly, the bias induced by this simple-yet-crude model for the shape of $O$ is smaller than when we used the full model, and a similar trend is observed for the standard deviation.
We conclude that although there is no a priori way to determine what parametric shape to use, the template procedure is not particularly sensitive to this choice.\footnote{In fact, we tried an even more radical example of assuming the shape of the parametric fit to $O$ was simply a line, and the closure test again works out surprisingly well.}

The fact that the template method does not require a model which accurately depicts the data can be seen as both a positive and a negative feature. 
On the positive side, it implies that one does not need to worry too much about the actual shape of the distribution when constructing a fit function, which is a plus since it is unknown how one might determine such shapes analytically (especially including the impact of pre-selection cuts).
On the other hand, this opens the possibility that physically unmotivated observables can be used, as long as they are relatively correlated with the top mass. 
The fact that a good fit is not a necessary requirement for closure in the template approach implies that subtle effects could bias the final extracted value of the top mass without warning when testing the self consistency of the procedure.
We will see an example of this kind of issue in \cref{sec:result}, where we investigate the impact of stealth stop contamination.

\clearpage
\pagebreak

\section{The Semi-leptonic Channel: A Modified Approach}
\label{sec:semi}
Although the main focus of this study is to quantitatively investigate the impact that light stops could have on the measurement of the top quark mass, in this section we will critically evaluate the application of the template method to the semi-leptonic final state as currently implemented by ATLAS in Ref.~\cite{Aaboud:2018zbu}.  
In particular, we showcase how defining a single observable for a pair produced event can obscure the physical interpretation of the observable.
This can be corrected by a straightforward implementation of a two-dimensional template.  
The issue and its resolution are presented in what follows.

\subsection{Pre-selection Cuts}
\label{sec:CutsSemiLep}
The defining characteristic of the semi-leptonic channel is that one of the top decays involves a lepton and the other decays fully hadronically.
The final state of interest is then two $b$-jets, two light flavor jets, one charged lepton, and missing energy from the neutrino.  
This is a powerful channel since the QCD background is reduced due to the lepton requirement.

We recast the ATLAS measurement~\cite{Aaboud:2018zbu} as closely as possible.  
However, we encountered a subtle issue as discussed in what follows, which motivates our modified approach.
The parton level events are generated using \texttt{Madgraph}~\cite{Alwall:2014hca}, and are subsequently showered and hadronized using \texttt{Pythia 8}~\cite{Sjostrand:2007gs}.  
We use \texttt{Delphes}~\cite{deFavereau:2013fsa} to simulate detector effects, and we modified the \texttt{Delphes} detector card to match the $b$-tagging characteristics reported by ATLAS. 
More details regarding the event generation can be found in \cref{sec:EventGeneration}, and additional details and validation results for the semi-leptonic channel are given in \cref{sec:semi2}.

In reconstructing objects for the analysis, we use the following set of definitions. 
Electron candidates are required to have a transverse momentum of $p_T >25$~GeV, $|\eta| < 2.47$ excluding the range (1.37, 1.52) due to the mismatch between the barrel and the end cap at ATLAS. 
Muon candidates must satisfy $p_T >25$~GeV and $|\eta| < 2.5$. 
Jet candidates are reconstructed with the anti-$k_t$ algorithm~\cite{Cacciari:2008gp} with a radius of $R=0.4$ and are required to satisfy $p_T >25$~GeV and $|\eta| < 2.5$. 
Muons reconstructed within $\Delta R < 0.4$ of a jet candidate are considered to be part of the jet and are subsequently removed from the list of charged lepton candidates. 
Jet candidates are labeled as jets if they have $\Delta R > 0.2$ from all electron candidates, and otherwise they are removed.  
Finally, electron candidates within $\Delta R < 0.4$ of a valid jet are removed. 
We set a flat $b$-tagging efficiency of 0.7, and rejection factors of 5 and 140 for the charm quark and the light quarks, respectively.

\clearpage
\pagebreak
To select events that are likely due to the semi-leptonic decay of a $t\bar{t}$ pair, the following pre-selection cuts are imposed:
\begin{itemize}
\setlength\itemsep{0em}
\item Exactly one charged lepton.
\item The \MET and $m_T^W$ cuts depend on the type of lepton:\footnote{$m_T^W$ is the transverse mass of the $W$ and is defined as $m_T^W = \sqrt{2 p_{T,\ell} ~\slashed{E}_T \Big(1-\cos \phi \big(\ell, \vec{\slashed{E}}_T \big) \Big)}$\,. } 
\begin{itemize}[label=$\bullet$]
\setlength\itemsep{0.5em}
\item $\mu$ channel: $\MET > 20\gev$ and $\MET +m_T^W > 60\gev$.
\item $e$ channel: $\MET > 30\gev$ and $m_T^W > 30 \gev$.
\end{itemize}
\item At least four jets with $p_T > 25 \gev$ and $|\eta| < 2.5$.
\item Exactly two $b$-tagged jets.
\end{itemize}
Table~\ref{table:cutflowsemi} in \cref{sec:semi2} shows the number of events that survive each of these successive cuts as predicted by our simulation.

\subsection{A Likelihood Approach to Inferring the Neutrino Momentum}
\label{sec:LikelihoodForPNeutrino}
A likelihood-based method is used to determine the missing neutrino momentum and address the combinatoric backgrounds as developed in Ref.~\cite{Erdmann:2013rxa}.
This methodology is the basis of the template approach as a function of the reconstructed top quark mass, $m_{t,\text{reco}}$, developed by ATLAS in the semi-leptonic channel as discussed in the next section.  
In order to recast this method, a likelihood function is built from Breit-Wigner (BW) distributions~\cite{Breit:1936zzb} defined as follows for each event that passes the preselection cuts:
\begin{equation}
\text{BW}\big(m \,\big|\, p\big) = \frac{1}{\big(p^2 - m^2\big)^2 + m^2\, \Gamma^2}\,,
\end{equation}
where $m$ is the particle mass and $\Gamma$ is its width.
%
%
%
The likelihood function is simply the product of four BWs, one for each of the two $W$ bosons, one for each of the two top quarks:
\begin{align}
\hspace{-10pt} L \big(\mtreco, m_{W,\rm{reco}}, p_{z,\nu} \,\big|\, p_{b_1}, p_{b_2}, p_{q_1}, p_{q_2}, p_{\ell}, \MET\! \big) =\hspace{3pt} &\text{BW}\big(\mtreco \,\big|\, p_{b_1} + p_{q_1} + p_{q_2}\big) \times  \nonumber\\[3pt]
& \text{BW}\big(m_{W, \rm{reco}} \,\big|\, p_{q_1} + p_{q_2}\big) \times \nonumber\\[3pt]
& \text{BW}\big(\mtreco \,\big|\, p_{b_2} + p_{\ell} + \MET + p_{z,\nu}\big) \times\nonumber\\[3pt]
& \text{BW}\big(m_{W, \rm{reco}} \,\big|\,  p_{\ell} + \MET + p_{z,\nu}\big)\,,
\label{eqn:like}
\end{align}
where $p_{b_{1,2}}$ are the four momenta of the two $b$-jets, $p_{q_{1,2}}$ are those of the untagged jets, and $p_{\ell}$ is the lepton four momentum.
ATLAS additionally includes transfer functions in the likelihood, $W(p_{\rm{measured}} | p_{\rm{true}})$, for each jet, the lepton, and the missing energy.  These transfer functions parameterize the mapping between the momenta of the detector-level objects and the momenta of the initial partons.\footnote{Note that these transfer functions are distinct from the jet energy scale.} The effect of the transfer functions will be discussed in more detail below.

The inputs to \cref{eqn:like} are the lepton momentum, the missing transverse momentum, and the momenta for up to six jets. 
The $x$ and $y$ components of the neutrino momentum are assumed to be equal to the missing energy components. 
The $z$ component, $p_{z,\nu}$ is unmeasurable at the LHC, and is therefore treated as a free parameter when maximizing the likelihood function, where the initial value provided to the maximizer is derived from $m_W^2 = \left(p_\ell + p_\nu\right)^2$.  
If the solutions of $p_{z,\nu}$ are complex, then the initial guess for the maximization is set to $p_{z,\nu}^\text{init} = 0$. 
If there are two real solutions, then the solution resulting in the largest likelihood is used.
The likelihood is then maximized for all possible assignments of the $b$-tagged jets to the leptonic side of the event, and all choices of two out of the possible four un-tagged jets. 
The choice which maximizes the likelihood is then taken to determine the assignment of decay products for both hadronic and leptonic tops. 
We have additionally checked that this approach does a reasonable job of reproducing the truth level assignments of final states with the appropriate top, and that it tends to find a very good approximation for the $z$-component of the neutrino momentum, as expected.

\subsection{The ATLAS Semi-leptonic Template}
\label{sec:like}
After selecting events using the preselection cuts described above in \cref{sec:CutsSemiLep}, ATLAS applies the likelihood method introduced in \cref{sec:LikelihoodForPNeutrino}.  
This provides a systematic way of assigning final state objects to either of the two top candidates, which is then used to construct a three dimensional template as a function of $m_t$, the jet energy scale (JES), and the $b$-JES.  
This is done by fitting to three observables $O$, $\mtreco, m_{W,\rm{reco}}$, and $R_{bq}$, where
\begin{equation}
R_{bq}=\frac{p_T^{b,\text{had}}+p_T^{b,\text{lep}}}{p_T^{q_1}+p_T^{q_2}}\,,
\label{eq:Rbq}
\end{equation}
where $p_T^{b,\text{had}}$ and $p_T^{b,\text{lep}}$ are the momenta of the $b$-jets assigned to the hadronic and leptonic sides of the even respectively, and $q_{1,2}$ are the light flavor jets that are associated with the decay of the $W$.

ATLAS finds that $m_{W,\rm{reco}}$ largely constrains the JES, while $R_{bq}$ constrains the $b$-JES relative to the JES. 
Given that our analysis relies on a simple parametrized detector simulation, we are not equipped to perform a realistic study of the impact of varying the JES or $b$-JES.  
Critically, we find that our $m_{W,\rm{reco}}$ and $R_{bq}$ distributions agree relatively well with those provided by ATLAS, as shown in \cref{sec:semi2} in the left and center panels of \cref{fig:semimatch}.  
Therefore, we are confidant that the JES and $b$-JES dependence will not have a significant impact on our interpretation of the semi-leptonic mass measurement.

The extraction of the Monte Carlo top quark mass comes mostly from the \mtreco distribution. 
ATLAS finds that \mtreco peaks at values much lower than the top mass that is extracted.
We find that we cannot reproduce the ATLAS distribution, in part because we neglect the transfer functions in the likelihood and ATLAS additionally uses a BDT to select events.
This is shown in  \cref{sec:semi2} in the right panel of \cref{fig:semimatch}.

\subsection{Impact of the Transfer Functions}
\label{sec:mod}
It is surprising that ATLAS finds that the distribution for $\mtreco$ peaks below the actual top mass.
%
It is important to emphasize that the location of the peak for ATLAS \emph{is not} the extracted value, which comes from finding the \emph{best-fit template}.  
We reiterate that as shown in Sec.~\ref{sec:tpl}, as long as the shape of the template varies linearly with the generator mass, the template procedure will close and the extraction of the best fit is expected to be robust. 
Despite this fact, this section is devoted to explaining the unexpected \mtreco distribution.  
Along the way, we will argue that \mtreco is not physically meaningful, which will motivate a physics-driven proposal for a modified approach presented in \cref{sec:2dmass}.

In order to generate their distribution, ATLAS populates a histogram using the value of \mtreco that maximizes the likelihood function in \cref{eqn:like} for each event.  
The underlying assumption is that \mtreco captures the best fit top mass for the whole event.  
However, this is not always the appropriate interpretation of this variable, as can be made clear by studying the form of the likelihood function without the transfer functions, as we do now.

As discussed above, the likelihood-based approach provides a way to systematically assign the final state objects to the two top quarks in the event, while also solving for the $z$-component of the neutrino momentum.
Assuming one has made all of these choices such that the maximum value for the likelihood can be achieved, we are left with a simple function of \mtreco\!:
\begin{equation}
\hspace{-5pt} L\sim \frac{1}{\Big[ \Big(\sum p^{\rm{had}}\Big)^{2}-\mtreco^{2} \Big]^{2}+\mtreco^{2}\, \Gamma^{2}}
\times
\frac{1}{\Big[\Big(\sum p^{\rm{lep}}\Big)^{2}-\mtreco^{2} \Big]^{2}+\mtreco^{2}\, \Gamma^{2}}\,,
\label{eq:Lmtreco}
\end{equation}
where $\sum p^{\rm{lep}}$ and $\sum p^{\rm{had}}$ are the sum of the four momenta for the final states assigned to the leptonic and hadronic tops respectively, and the width $\Gamma$ is set to the PDG value, 1.41 GeV.

\begin{figure}[t]
\begin{center}
\includegraphics[width=0.85\linewidth]{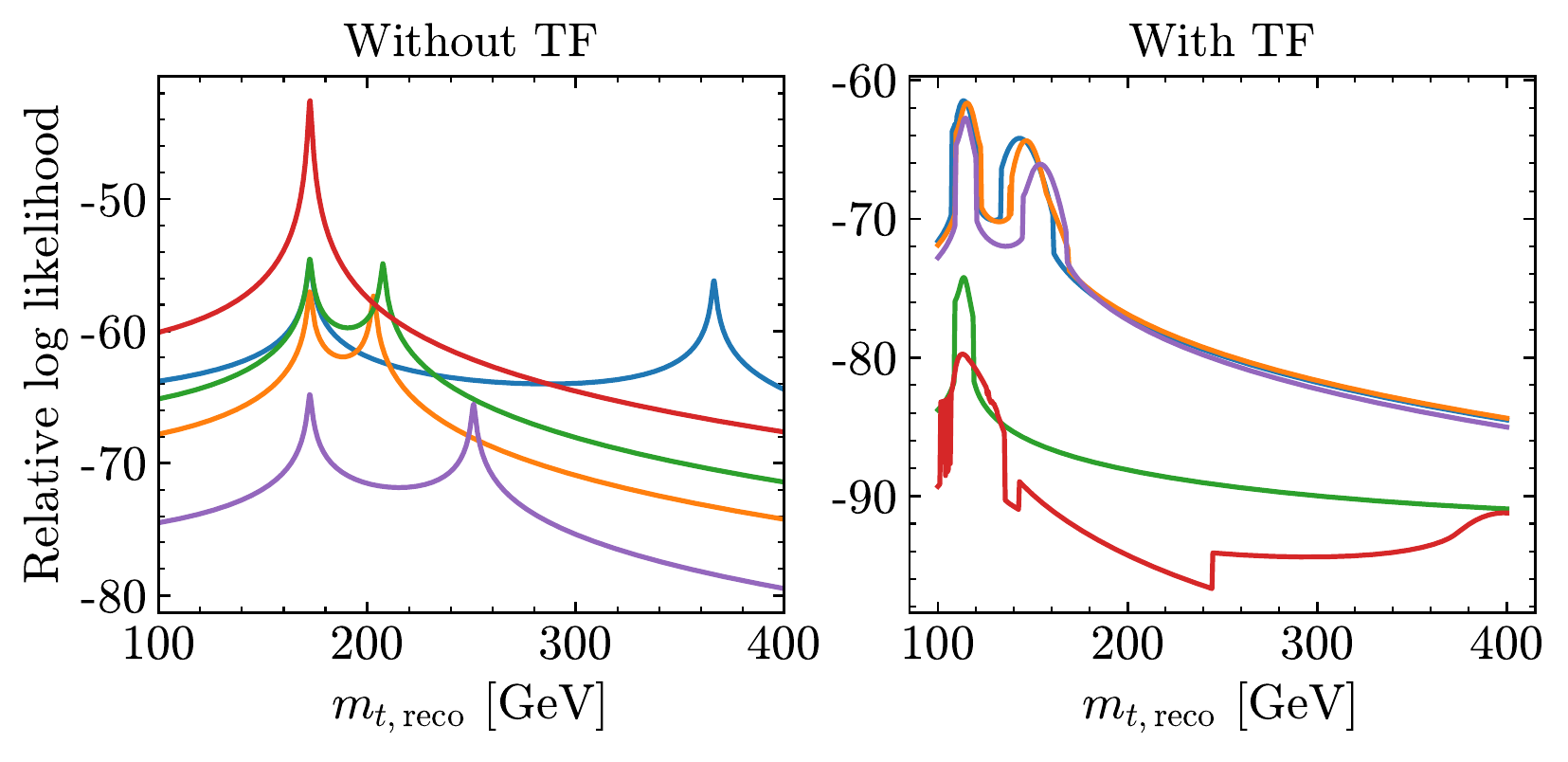}%
\caption{
The left panel shows double peak structure of the log likelihood function for 5 random events at truth level, without including the transfer functions. 
A clear double peaked structure occurs for many events.
As expected, the lower mass peak is always more likely than the higher mass one.
The right panel shows the impact of including the transfer functions in the likelihood and optimizing over the $z$-component of the neutrino momentum for the same five events, as implemented in \textsc{KLFitter}.
Although the transfer functions push the peaks together, they do not always result in a likelihood that has a single local maximum.
Note that the absolute scale of the log likelihood is expected to differ between the two panels.
}
\label{fig:doubpk}
\end{center}
\end{figure}

The choice to use a Breit-Wigner shape when constructing the likelihood function that peaks at the best fit mass of the top quark is clearly physically motivated. 
However, while the product form in \cref{eqn:like} works very well as an approach to the combinatoric background and for determining $p_z$ for the neutrino, it does not return an event-level ``best fit'' for the top mass.  
In particular, using the simplified expression in \cref{eq:Lmtreco}, it is straightforward to see that the likelihood shape has two very sharp peaks, one for each choice of $\mtreco^2$ that equals $\big(\sum p^{\rm{had}}\big)^{2}$ and $\big(\sum p^{\rm{lep}}\big)^{2}$, rather than one peak between the two.
This point is clearly illustrated in the left panel of \cref{fig:doubpk}, where we evaluate \cref{eq:Lmtreco} as a function of \mtreco for five independent top pair production events (without the transfer functions) taken from the \textsc{KLFitter} semi-leptonic example file.\footnote{\href{https://github.com/KLFitter/KLFitter}{https://github.com/KLFitter/KLFitter}}  
%
%
Note that the left peak will always be more likely than the right one; this is clear from the denominator of the Breit-Wigner.\footnote{ATLAS states that they are using a ``Breit-Wigner'' distribution, but they do not specify its functional form.  If they had chosen to include the numerator factor $$k=\frac{2\, \sqrt{2}\,m\,\Gamma \sqrt{m^2\,\big(m^2+\Gamma^2\big)}}{\pi\, \sqrt{m^2+\sqrt{m^2\big(m^2+\Gamma^2\big)}}}\,,$$ as the distribution is sometimes given, the logic changes and the right peak is always more likely.  We note that the \texttt{ROOT} implementation~\cite{ROOTfitting} of the Breit-Wigner function has unit numerator.} 

The transfer functions are implemented in part to help account for the two peaks, by having the ability to adjust the jet momenta to push the two tops to the same $p^2$. 
There are two minor issues with this.
The first is that off-shell effects are relevant, and even at the partonic, truth-level, the two tops rarely have the same $p^2$. 
The second is related to the fact that the Breit-Wigner function has higher likelihoods for lower invariant mass peaks, which allows the transfer functions to consistently push \mtreco to smaller values.
The right panel of \cref{fig:doubpk} shows the likelihood when using the transfer functions for the same events as the left panel.
The sharp double peaks are softened---and sometimes combined into one---but the peaks are also shifted to lower masses.


The ``best-fit'' momenta maximizing the likelihood are not physical. 
The procedure results in \mtreco well below the actual top mass; because of this, ATLAS uses the measured momenta for $m_{W,\rm{reco}}$.
Part of the motivation for the transfer functions was to combine the two tops into a single observable, at the expense of reducing the available information.
Without the transfer functions, we find the value of \mtreco that maximizes the likelihood corresponds to the peak associated with the leptonic top $\sim 60\%$ of the time.  
This implies that the \mtreco distribution is a non-trivial mixture of hadronic and leptonic tops, with unknown implications for systematic effects on the distribution of \mtreco\!\!.  
%
%
%
This motivates our proposal for a modified approach using independent information from both tops, which is presented in the next section.

\subsection{A Two-dimensional Mass Extraction Template}
\label{sec:2dmass}
\begin{figure}[t]
\begin{center}
\includegraphics[width=0.5\linewidth]{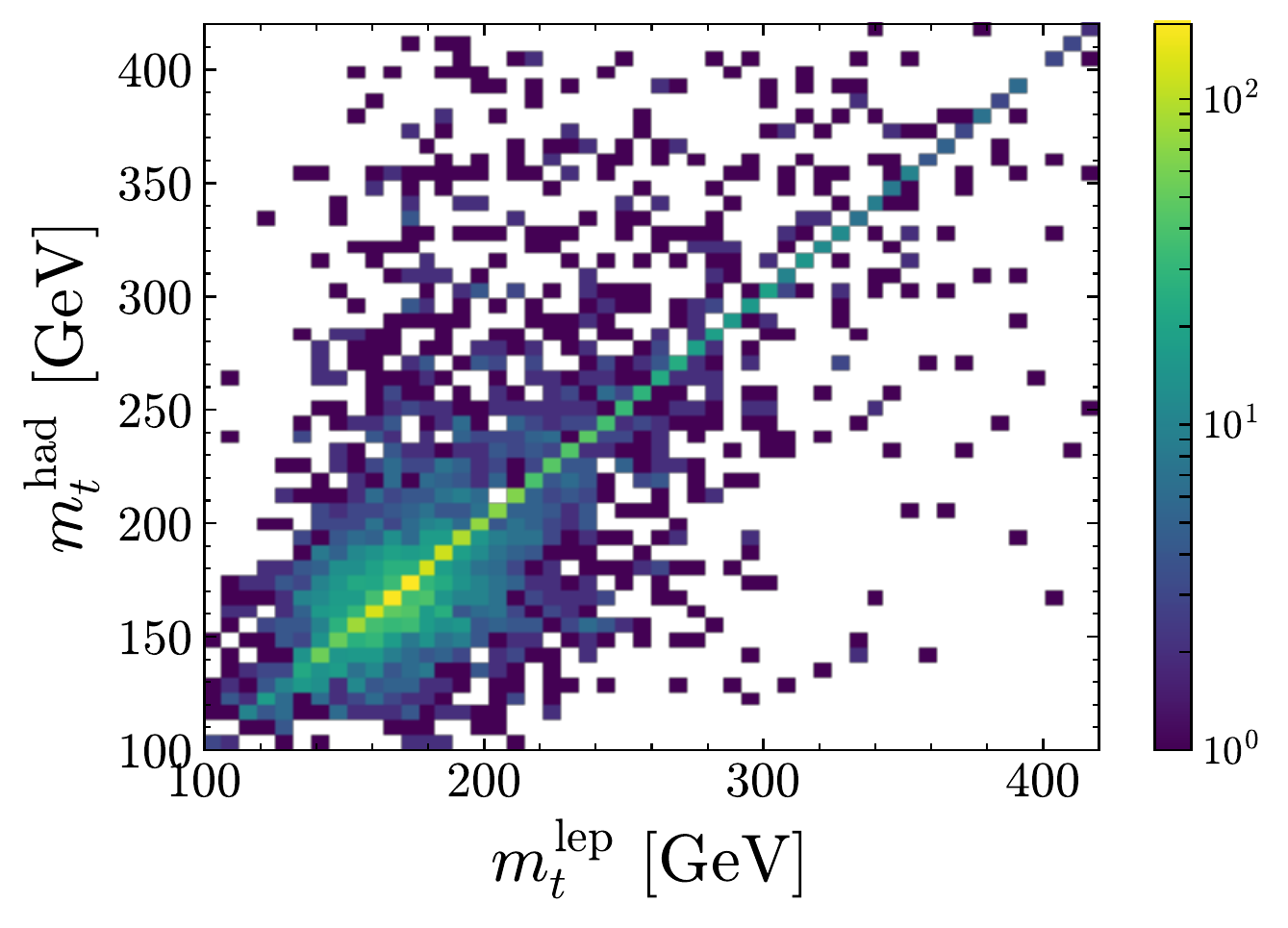}%
\caption{This figure provides a density plot of the reconstructed top mass for the semi-leptonic channel in the hadronic mass versus the leptonic mass plane.  For the underlying simulations, the Monte Carlo truth mass is taken to be $m_t$=172.5 GeV.  In the majority of events, the two masses are highly correlated and lie along the diagonal centered around the truth value of $m_t$.  This shape motivates the form of the fitting function we are proposing that can be used to build a template for  extracting the top mass in the semi-leptonic channel.}
\label{fig:2dcorr}
\end{center}
\end{figure}

Instead of using a one-dimensional template for \mtreco\!, one would prefer an approach that takes advantage of the fact that there is both a leptonic and hadronic top decay in each event. 
We propose a modified approach in this section, relying on the same combined likelihood given in \cref{eqn:like} to control the combinatorics and to solve for the missing neutrino momentum. 
We use the configuration that maximizes the likelihood to generate the distribution shown in \cref{fig:2dcorr}, where we give a two-dimensional density plot of the hadronic and leptonic top masses that result. 
One observes that the density is essentially symmetric about the diagonal $m_{t}^{\rm{had}} = m_{t}^{\rm{lep}}$, and that most of the time the values of the top masses from the two sides of the event are very similar. 
Along the diagonal, the density peaks near $m_{t} \sim 170\gev$ and then has an extended tail to larger masses. 
This 2D plane provides an excellent candidate for an improved observable $O$ from which we construct a template.

For our parametric model, we want a function with a peak and a tail along the diagonal. 
We chose this to be a Gaussian plus a Landau function, following the ATLAS approach used to fit a one-dimensional \mtreco distribution. 
Then we model the spread orthogonal to the diagonal using a second independent Gaussian. 
Concretely, the two-dimensional template is 
\begin{align}
P\big(x, y; a, b, \chi_1, \sigma_1, \mu, c,  \chi_2, \sigma_2 \big) =&\hspace{3pt} 
\Big(a\,P_{\rm{\,Gaussian}}\big(x; \chi_1, \sigma_1\big) + b\,P_{\,\rm{Landau}}\big(x;\mu, c\big) \Big) \nonumber \\[4pt]
&\hspace{10pt}\times P_{\rm{\,Gaussian}}\big(y; \chi_2, \sigma_2\big)\,,
\label{eqn:2dtemplate}
\end{align}
where $x$ and $y$ are the distance along the diagonal and distance away from the diagonal, respectively. 
This is a relatively crude model for the distribution shown in \cref{fig:2dcorr}, and it does not take into account how the spread away from the diagonal changes as a function of the distance from the origin. 
We tested that a more precise fitting function did not lead to improved extractions of the Monte Carlo top mass, while drastically increasing the computational time to perform the two-dimensional fit.
This makes sense given the discussion regarding the sensitivity of the template approach to the shape of the fit function, as discussed in \cref{sec:choice}.

As in our toy model, we perform a closure test to validate the proposal of extracting the top mass from the two-dimensional template. Our new approach faithfully extracts the correct mass, and comes with a relatively small statistical uncertainty $\sim 0.1 \gev$. 
We additionally checked that varying the linear fit of the templates up and down by an amount determined by the covariances led to a similar size uncertainty; see~\cref{sec:toy} for a discussion of this test. 
Our determination of these sources of uncertainty due to the template method are subdominant to the JES uncertainties provided by ATLAS in~\cite{Aaboud:2018zbu}.

With this modified procedure in hand, we are now ready to assess the impact of stop contamination on the top mass measurement.
As we will emphasize below, our results in the semi-leptonic channel use the two-dimensional template method discussed here.
As such, the results in the semi-leptonic channel are \emph{not} a recasting, but can instead be interpreted as an estimate for how much contamination one could expect for this final state.

\section{The Impact of Light Stops}
\label{sec:result}

Now that we have explored the template method as it is used by ATLAS to extract the Monte Carlo mass of the top quark (along with our modified approach in the semi-leptonic channel), we will turn to the impact of stop contamination on the template mass extraction. 
This is important since attempts by ATLAS~\cite{Aad:2014kva, Aad:2014mfk, Aaboud:2019hwz} to exclude the stealth stop region of parameter space utilizing properties of high purity $t\bar{t}$ samples assume the top mass is measured in an orthogonal channel.
As we will show in this section, light stops can bias the extracted top mass by up to $2$~GeV.
This implies that any limit which claims to exclude stealth stops using aspects of the top pair kinematics must simultaneously account for the impact on the top mass measurement.
While this may not seem like a major issue at first glance, we emphasize that the leading order cross section prediction for $t\bar{t}$ production at $\sqrt{s}=8\tev$ drops from 160~pb for $m_t = 172 \gev$ to 150~pb for $m_t = 174 \gev$.
Since the production of stealth stops is  $\mathcal{O}(10 \pb)$, this could easily impact the boundaries of exclusion regions. This sensitivity to $m_t$ has been demonstrated by ATLAS in Ref.~\cite{Aad:2015pfx} where observed limits on the stop mass drop from around 180 GeV to 160 GeV if the top mass is changed from 172.5 to 175 GeV.
Here, we will focus on demonstrating the quantitative impact of this contamination --- assessing how this alters limits is left for future work.

For concreteness, we will work with the stop-neutralino Simplified Model framework.  
Given a choice of top mass, we then generate a suite of events for different values of the stop mass (and for two benchmark choices of the neutralino mass), including the full effects of the off-shell propagators following the procedure detailed in Ref.~\cite{Cohen:2018arg}; more details regarding the event generation can also be found in \cref{sec:gen}.
In particular, this approach self consistently computes the width of the stop and the top quark as the parameter space is varied.
The pair production of stops is determined by its QCD interactions, and its subsequent decay $\stopone \rightarrow t^{(*)}~\ninoone$ yields a (potentially off-shell) top quark and missing energy.
Intuitively, the biggest impact on the top mass measurement will occur in the parameter space where the top that results from the stop decay is off-shell, since the reconstructed ``top'' in such events will have a ``mass'' that is smaller than $m_t$.
We will see exactly this behavior in the quantitative results that follow.

\begin{figure}[t]
\begin{center}
\includegraphics[width=0.98\textwidth]{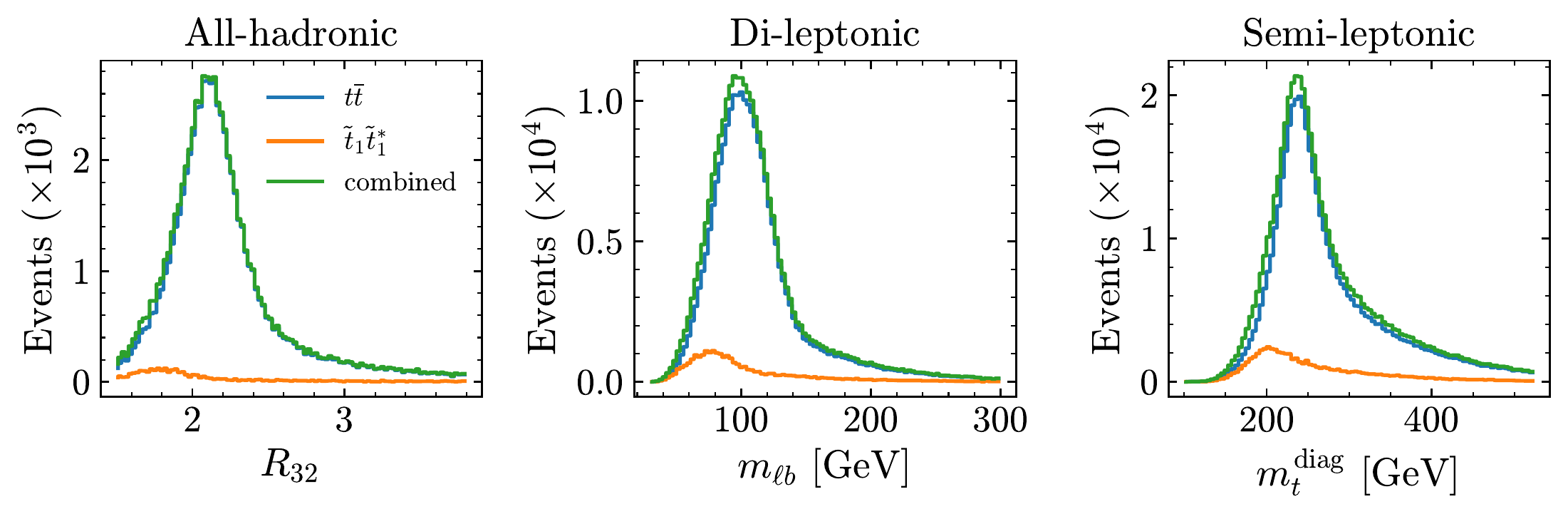}
\caption{
The distributions that are used as input to the template procedure.
Moving from left to right shows all three channels for the $t \bar{t}$ sample (blue), the $\tilde{t}_{1} \tilde{t}_{1}^{*}$ sample (orange), and the combination (green), where the masses are taken to be $m_t =$ 172.5 GeV and $\mstop =$ 164 GeV with $\mnino =$ 1 GeV. 
Stop contamination tends to shift the distributions slightly to the left.  
}
\label{fig:example}
\end{center}
\end{figure}

To get a sense of the impact that stealth stops can yield, \cref{fig:example} shows the shape of the potential stop contribution to the observable $O$ used to generate the template for each channel from top pair production with $m_t = 172.5 \gev$ (blue solid), stop pair production (orange solid) with $\mstop = 164 \gev$, $\mnino=1\gev$, and the combined distribution (green solid).
Note that in the semi-leptonic channel, we use the two-dimensional observable introduced in \cref{sec:2dmass}, but plot the one-dimensional slice along the $m_t^\text{diag} \equiv m_t^\text{had} = m_t^\text{lep}$ diagonal.
Each of these distributions are normalized using the production cross section times efficiency to pass the relevant pre-selection cuts, assuming an integrated luminosity of ${\cal L} = 20.2$ fb$^{-1}$. 
While the stop contribution is clearly subdominant, it peaks at a slightly lower value in each observable than $t\bar{t}$. 
This has the effect of biasing the combined sample such that the extracted Monte Carlo top mass that best fits the combined distribution is lower than the true value of $m_t$.

\afterpage{\clearpage}
\begin{figure}[p] 
\begin{center}
\vspace{-30pt}
\includegraphics[width=0.92\linewidth]{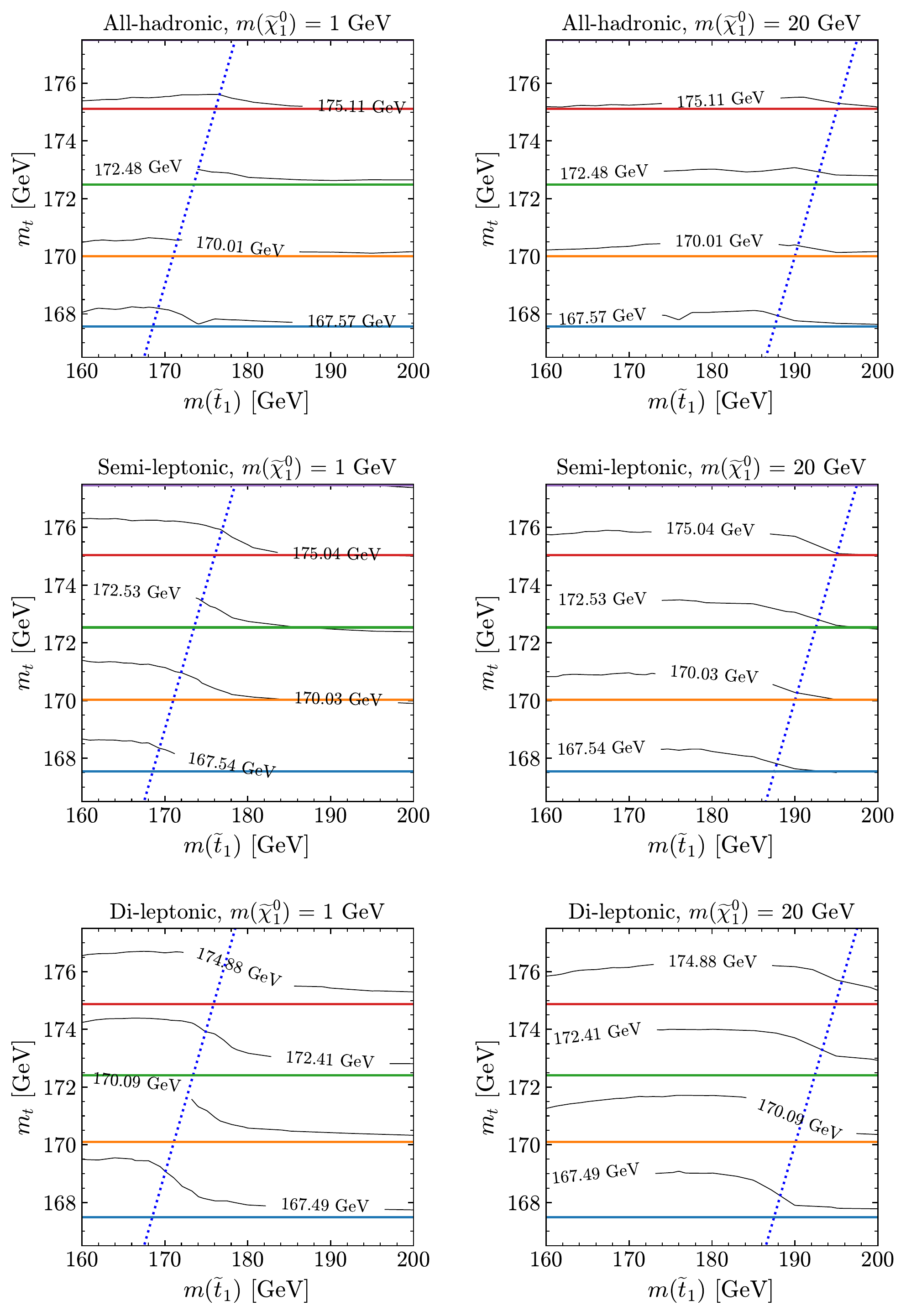}%
\caption{\small Effects of stop contamination on $m_t$ compared to pure top simulations. The top, middle, and bottom rows correspond to the all-hadronic, semi-leptonic, and di-leptonic channels, respectively. The plots in the left column correspond to the choice $\mnino=1\gev$, while those in the right column are for $\mnino = 20\gev$. In each panel, the colored lines represent the $m_t$ reconstructed from pure $t\bar{t}$ benchmark samples. The blue dotted line indicates the kinematic boundary where the tops from the stop decays start to be produced on-shell. The black curves are contours of constant reconstructed top mass when using SM only templates. The specific contours are chosen to match the SM only closure tests. 
} 
\label{fig:contam1}
\end{center}
\end{figure}

\begin{figure}[p] 
\begin{center}
\vspace{-30pt}
\includegraphics[width=0.92\linewidth]{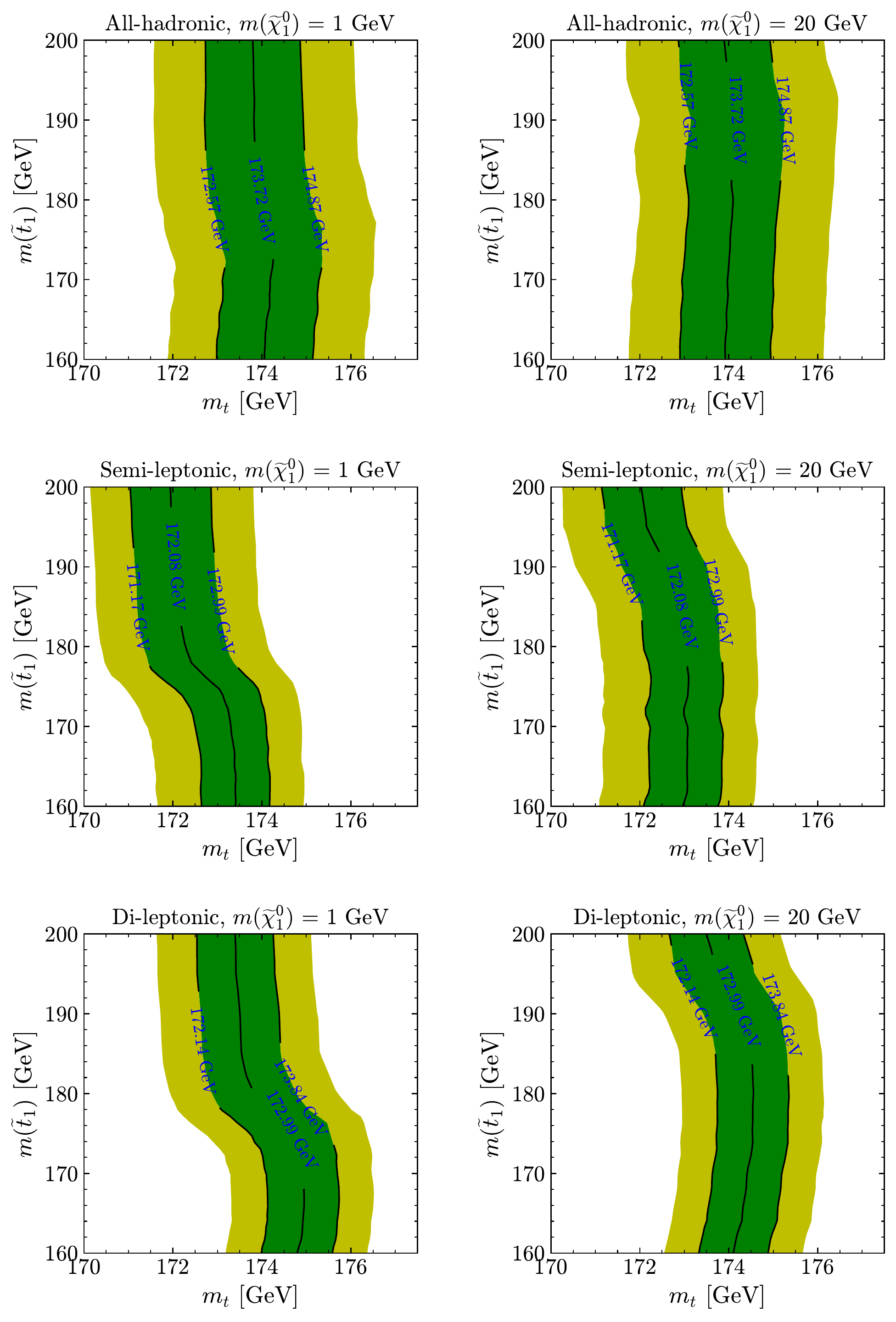}%
\caption{\small Contours of the mass extracted when using templates constructed in the SM only assumption. The top, middle, and bottom rows correspond to the all-hadronic, semi-leptonic, and di-leptonic channels, respectively. The plots in the left column correspond to the choice $\mnino=1\gev$, while those in the right column are for $\mnino = 20\gev$. In each panel, the central line denotes the value measured by ATLAS in the given channel. The green (yellow) band illustrates the regions in parameter space where the reconstructed mass is within 1-$\sigma$ (2-$\sigma$) of the measurement using the reported uncertainty. 
}
\label{fig:contam2}
\end{center}
\end{figure}

The results of our study for all three channels are presented in two different ways: the first representation is provided in \cref{fig:contam1}, and the second is in \cref{fig:contam2}.\footnote{Both plots show contours of the mass which would be extracted from templates  made using the SM only assumption when the real data is contaminated by stops. The contours are made in the same way for both plots, but different contours are shown to highlight different aspects of the contamination.}
The colored horizontal lines in \cref{fig:contam1} denote the mass that is extracted using the pure SM sample for true top masses of 167.5, 170, 172.5, and 175 GeV, respectively, as one moves from the bottom to top of each panel.
The thick black lines show contours of constant reconstructed top mass when using the templates made using only $t\bar{t}$.
Values of the contours are chosen to match the closure test values for the pure $t\bar{t}$ samples.
The dotted diagonal line shows the kinematic boundary where $\mstop = \mnino + \mt$.
Left of this dotted kinematic boundary, the tops are off-shell so the black lines are above the horizontal benchmark lines.
This implies that the truth-level top mass (shown on the y-axis) is larger than the reconstructed value when using a SM only template.
As the stops are taken to be heavier and cross the dotted line, two effects become important: the stops decay to on-shell tops removing the off-shell effects, and the stop production cross section decreases, thus explaining why the results asymptote to the pure SM in this limit.

The top row of \cref{fig:contam1} shows the results for the all-hadronic channel (for more details on the recast procedure for this channel see App.~\ref{sec:had}). 
For $t\bar{t}$ production, this channel does not result in any intrinsic missing energy, and the preselection cuts do not make any requirements on $\MET$\!.
This implies that the distribution utilized for the all-hadronic channel is less sensitive to presence of the additional $\MET$\! due to the final state neutralinos. 
Therefore, this channel is relatively insensitive to stop contamination; off-shell effects (left of the dotted line in the top left panel of \cref{fig:contam1}) yield the dominant impact on the top mass extraction.

The middle row of \cref{fig:contam1} shows the results for the semi-leptonic channel. 
Due to our issues validating this channel as discussed in~\cref{sec:semi} above, we have performed this analysis using our proposed 2D template approach. 
As with the all-hadronic case, when the stops are lighter (to the left of the blue dashed line) they decay through an off-shell top quark, which biases the templates to extract lower masses.
However, there is an additional important effect, which makes the results in this channel even more striking.
The SM contribution contains a neutrino, and so the pre-selection cuts explicitly rely on \MET\!.
Furthermore, the likelihood procedure utilized for addressing the combinatoric background and the missing $z$-component of the neutrino momentum assumes that the measured \MET corresponds to the transverse components of the neutrino momentum.
This implies that the neutralinos in the final state will have a non-trivial impact on the shape of the observable used for the template method. 
From the figures, it is clear that the impact of stealth stop contamination on the semi-leptonic channel is more dramatic than in the all-hadronic channel, yielding a bias as large as $\sim 2 \gev$.
For the largest stop masses, the reconstructed top mass over-shoots the true value in this range due to the effect of the neutralinos on the observable, but eventually asymptotes to the SM-only true value. 

\afterpage{\clearpage}

The di-leptonic channel is shown in the bottom row of \cref{fig:contam1}.
In this case, the SM final state contains two neutrinos, and so the preselection includes a cut on \MET.
As opposed to the semi-leptonic case, the observable used in this channel is simply the invariant mass of the lepton and $b$-jet pairs $m_{\ell b}$, and so the distribution should not be impacted by new sources of \MET.  
However, $m_{\ell b}$ it is not fixed by the mass of a parent particle, and so the resulting distribution is more sensitive to details such as the spin of the top quarks and the kinematics of the top pairs.
This explains why the bias in the reconstructed top mass for this channel is the most dramatic of the three, including the fact that the result asymptotes to the SM value even more slowly as the stop mass increases.

Now that we have a sense of how large the bias from stealth stop contamination can be, \cref{fig:contam2} illustrates the consistency of the BSM parameter space with the observations performed by ATLAS.
In this figure, the axes have been rotated with respect to \cref{fig:contam1}, and we plot the \emph{truth-level} Monte Carlo top mass used to generate events along the horizontal axis, while the input stop mass is on the vertical axis. 
For each point in the truth parameter space, we extract the reconstructed top mass. 
The black line at the center of the bands denotes the parameters that yield a reconstructed top mass which is equal to the value observed by ATLAS in each channel assuming the SM alone, while the green (yellow) bands are the 1-$\sigma$ (2-$\sigma$) uncertainties taken directly from the ATLAS papers~\cite{Aaboud:2017mae,Aaboud:2016igd,Aaboud:2018zbu}. 
This allows one to visualize the non-trivial shapes that result from stealth stop contamination in each channel, and provides some insight into what parameter choices could yield the best consistency.

\begin{figure}[t!]
\center
\includegraphics[width=0.85\linewidth]{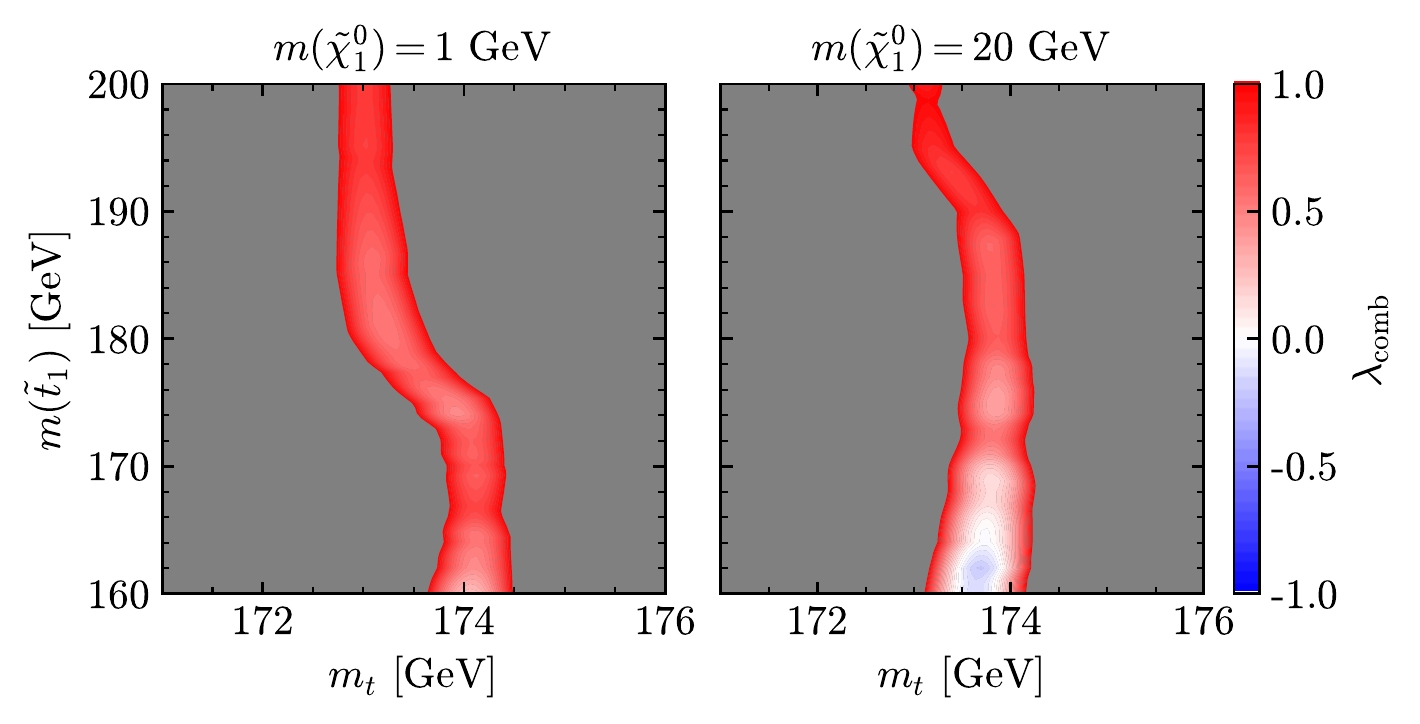}
\caption{
Values of the test statistic given in \cref{eqn:Lambda}, which is designed to compare the goodness of fit for the SM alone versus the BSM scenario studied here.
The red regions provide a fit which is within $1\sigma$ of the SM alone hypothesis, the white regions give the same quality of fit as the SM, and the blue regions give a better fit than the SM alone. 
The overall best fit is at $\mt = 173.7 \gev$ and $\mstop=162.0\gev$ for the panel with $\mnino = 20 \gev$; the fit to the top mass measurement at this point is only $0.1\sigma$ better than can be achieved with the SM only hypothesis.
}
\label{Fig:BSM_to_SM}
\end{figure}

To quantitatively explore the consistency between the three channels, we performed a naive combination of these channels in the BSM parameter space; the methodology is described in \cref{Sec:Combined} and the results are provided in the middle and right panels of \cref{fig:SMCombination}.
At each point in the $\mt$ - $\mstop$ plane (keeping $\mnino$ fixed), we compute the $\chi^2$ for the extracted template mass in each channel as compared to the observations. 
The orange stars in the middle and right panels of \cref{fig:SMCombination} show the best-fit point, and the shaded area shows the $1$-$\sigma$ region.
For the model with $\mnino = 1\gev$, the best fit point is found to be $\mt = 174.0 \gev$ and $\mstop=160\gev$,\footnote{Note that the best fit occurs at the lower edge of the $\mstop$ range simulated here; to find the true best fit, the region should be extended. However, the purpose of this study is to show the effect of stop contamination, which is clearly demonstrated, as opposed to taking any particular improvement in the fit seriously.} 
and when $\mnino=20\gev$, the best fit point is at $\mt=173.7\gev$ and $\mstop = 162.0\gev$. 
Both panels show that the data fit best using lighter stops; the 1-$\sigma$ uncertainty band for the right panel does not even extend to the top of the panel.
In addition, the entire region results in masses larger than the SM-only assumption.
If there are light stops, we may not know the mass of the top quark as accurately as we think we do.

As an amusement, we note that the all-hadronic and semi-leptonic uncertainty  bands only slightly overlap in the SM alone assumption.
With that, it may be possible that light stops could improve the consistency of the experimental results.
In order to naively explore the extent to which the BSM model is a better fit than the SM alone, we compute the test statistic defined in \cref{eqn:Lambda} and the result is presented in \cref{Fig:BSM_to_SM}.
While there is no particular overall improvement in the fit for $\mnino = 1\gev$, the heavier choice $\mnino = 20\gev$, does has a very mild preference for the BSM scenario.
Although we simply take this to be a coincidence given the current state of the top mass measurement, it does demonstrate that if the top mass measurements became discrepant between the different channels, light stops could bias the mass measurements enough to provide a resolution.


\section{Conclusions}
\label{sec:disc}
In this work, we have investigated the stealth stop contamination of the $t \bar{t}$ sample that can potentially bias the measurements of the top mass at ATLAS by up to $2 \GeV$. 
Three decay channels are studied in detail: all-hadronic, di-leptonic, and semi-leptonic. 
The top mass measurement in the all-hadronic channel is the least sensitive to stop contamination, while the di-leptonic channel is the most sensitive. 
The combination of results suggests that the heavy neutralino case is slightly favored in terms of overall consistency among the three channels. 
Furthermore, we have proposed a modified method to better measure the top mass in the semi-leptonic channel.

While our focus here was on the template method, there are many approaches that could be used to measure the top mass, which could respond differently to stop contamination.
For example, CMS has published a measurement of $m_t$ using 5 fb$^{-1}$ of 7 TeV data utilizing the $m_{\ell b}$ endpoint in the di-leptonic channel~\cite{Chatrchyan:2013boa}.  
Since this is a kinematic feature (as opposed to relaying on the shape of the entire distribution) the relevant kink is unlikely to be modified by the presence of light stops.
It would be interesting to investigate how contamination would impact non-template based approaches like this one.

In conclusion, $\mathcal{O}(1\GeV)$ shifts in the top mass measurement due to stop contamination are possible and can have $\mathcal{O}(10\GeV)$ impacts on the stealth stop exclusion limits~\cite{Aad:2015pfx}. 
Thus, we advocate that the LHC experiments perform an analysis of the full three-dimensional Simplified Model parameter space spanned by $\mstop$-$\mt$-$\mnino$ in order to make a definitive statement on the potential existence of stealth stops.


\section*{Acknowledgments}

The authors thank Walter Hopkins for useful comments on the manuscript.
We are particularly grateful to Merve Karacasu, Andrea Knue, Richard Nisius, and Javier Pena for discussions on the semi-leptonic channel.
TC and PZ are supported by the U.S. Department of Energy under grant number DE-SC0018191. 
SM is supported by the U.S. Department of Energy under grant numbers DE-SC0012008 and DE-SC0017996.
BO is supported by the U.S. Department of Energy under grant numbers DE-SC0018191, DE-SC0011640, and DE-SC0013607. 
This work utilized the University of Oregon Talapas high-performance computing cluster.

\appendix
\section*{Appendices}
\addcontentsline{toc}{section}{\protect\numberline{}Appendices}%

\section{Event Generation}
\label{sec:EventGeneration}
 \subsection{Top Event Generation}
 \label{sec:gentops}
The 8 TeV $t \bar{t}$ sample is generated at the parton level using \texttt{MadGraph5\_aMC@NLO 2.6.1}~\cite{Alwall:2014hca}, and is passed to \texttt{Pythia 8.2}~\cite{Sjostrand:2007gs} for showering and hadronization. 
Detector effects are approximated using \texttt{Delphes 3.4.1}~\cite{deFavereau:2013fsa}, which relies on \texttt{Fastjet}~\cite{Cacciari:2011ma, Cacciari:2005hq} to cluster the jets with the anti-$k_T$ algorithm~\cite{Cacciari:2008gp}. 
We use the default \texttt{Delphes} ATLAS card, except that the $b$-tagging efficiency is set to be 0.57 for all-hadronic channel and 0.7 for the other two channels, in accordance with ATLAS~\cite{Aaboud:2017mae, Aaboud:2016igd, Aaboud:2018zbu}. 
We generated 5 million events for each of 5 top masses: 167.5 GeV, 170 GeV, 172.5 GeV, 175 GeV, 177.5 GeV.

\subsection{Stop Event Generation}
\label{sec:gen}
We work with a stop-neutralino Simplified Model, where the stop has the couplings appropriate for being right-handed. To cover the stealth stop region, events are generated for two choices of $\mnino$: 1~GeV and 20~GeV, and for a range of stop masses: $\mstop$ from 160~GeV to 180~GeV in steps of 2~GeV, and $\mstop$ from 180~GeV to 200~GeV in steps of 5~GeV.
At each parameter point, we use \texttt{MadGraph5\_aMC@NLO} to calculate the stop decay width. 
One must be very careful to account for all finite width effects during the generation of events when the top can be off-shell, see~\cite{Cohen:2018arg} for a detailed discussion. 
To this end, we ensure that the top and $W$ widths are defined consistently for the decay and production in \texttt{MadGraph5\_aMC@NLO}. 
Given the appropriate widths, we again use \texttt{MadGraph5\_aMC@NLO} to calculate the matrix elements and generate 500,000 events for stop production and subsequent decay to each final states. 
We emphasize that this approach does not require any particle to appear on shell, and keeps track of all spin correlations and finite width effects.

To mix the stop and top samples so that we can investigate the impact of the stop contamination, we weight the events from the two samples according to their leading order cross sections, appropriately normalized by the total number of events generated.  
The stop production cross section is approximately $O(10\%)$ of the top, when the stop mass is within the range we scan.

\section{The All-hadronic Channel}
\label{sec:had}
In the all-hadronic channel, the final states is characterized by two $b$-jets and four light-flavor jets. 
While this channel has the largest branching ratio (45.7$\%$) of the three final states, it suffers from a large QCD multi-jet background and from large uncertainties in the JES. 
This channel is the most challenging to measure, which explains why it has the largest error bar.
 \subsection{Pre-selection Cuts}
 \label{sec:cut1}
The following preselection cuts are required before applying the template procedures. 
Events with isolated $e/\mu$ are excluded. 
At least 6 jets with $p_T >$ 25 GeV and $ |\eta| < 2.5$ are required, and at least 5 of these jets must have $p_T >$ 60 GeV. 
For any pair of jets, an isolation requirement is applied such that $\Delta R(j_i, j_k) >$ 0.6, where $\Delta R$ is the angular distance between two objects. 
An event must contain at least 2 $b$-tagged jets, with an azimuthal separation of $\Delta \phi(b_i,b_j) >$ 1.5. 
To remove events with neutrinos, a missing transverse energy cut of $\MET <$ 60 GeV is applied.

\begin{figure}[t]
\begin{center}
\includegraphics[width=0.85\linewidth]{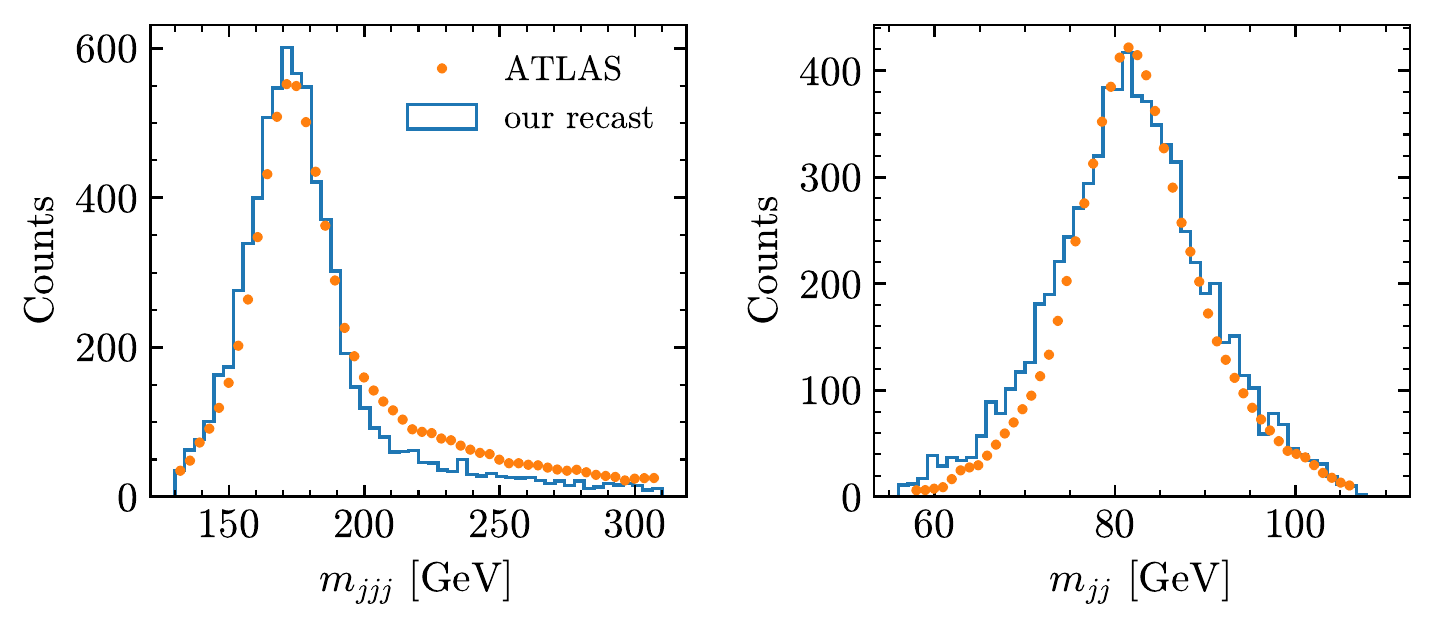}%

\caption{
Two validation plots are provided for the all-hadronic channel.  
Three-jet invariant mass distributions for top candidates (left) and di-jet invariant mass for $W$-boson candidates (right), assuming $m_t=172.5$ GeV. 
The blue distributions correspond to our simulation and the orange dots represent the ATLAS results~\cite{Aaboud:2017mae}, which have been rescaled to so that the normalizations agree.
} 
\label{fig:mjjvsmjjj}
\end{center}
\end{figure}

The all-hadronic channel has large combinatoric background, due to the homogeneity of the final state. 
To associate the jets with a particular top decay, a minimum $\chi^{2}$ approach is utilized.
One keeps the permutation that gives the lowest $\chi^{2}$ among all possible permutations of jets in an event, where the $\chi^{2}$ is defined as 
 \begin{equation} 
   \chi^2 = \frac{\big(m_{b_1 j_1 j_2} - m_{b_2 j_3 j_4}\big)^{2}}{\sigma_{\Delta m_{bjj}}^{2}} + 
                    \frac{\big(m_{j_1 j_2} - m_{\Wboson}^{\textrm{MC}}\big)^{2}}{\sigma_{m_{\Wboson}^{\textrm{MC}}}^{2}} +
                    \frac{\big(m_{j_{3} j_{4}} - m_{\Wboson}^{\textrm{MC}}\big)^{2}}{\sigma_{m_{\Wboson}^{\textrm{MC}}}^{2}}\,, 
 \label{eq:newchi2}
\end{equation}
where the $m_{\Wboson}^{\textrm{MC}}$ is taken to be $81.18 \pm 0.04$ GeV and the widths $\sigma_{\Delta m_{bjj}}$ and $\sigma_{m_{\Wboson}^{\textrm{MC}}}$ are taken from \cite{Aaboud:2017mae}: $\sigma_{\Delta m_{bjj}} = 21.60 \pm 0.16 \gev$ and $\sigma_{m_{\Wboson}^{\textrm{MC}}} = 7.89 \pm 0.05 \gev$. 
Then the preselection requires $\chi^{2}<$11.
Finally, a cut is applied to the azimuthal angle between $b$-jets and their associated $W$ boson: the average of the two angular separations between the $b$ and the $W$ for each event must satisfy $\langle \Delta \phi(b,\Wboson) \rangle <2$. 
For validation, we present \cref{fig:mjjvsmjjj}, which gives distributions for the three and two jet invariant masses, after the pre-selection cuts are applied.

\begin{table}[t]
\centering 
\renewcommand{\arraystretch}{1.3}
\setlength{\tabcolsep}{20 pt}
\begin{tabular}{l r} 
\toprule
Cut & \multicolumn{1}{c}{All-hadronic}\\[0.5ex] 
\hline 
Before cuts &  108433\\ 
$e/\mu$ isolation &  108367\\
$\geq$ 6 jets with $p_T >$ 25 GeV & 71173\\
$N_{b tag} \geq$ 2 & 23008\\
$\Delta R(j_i, j_k) >$ 0.6 & 18241\\
$\geq$ 5 jets with $p_T >$ 60 GeV & 4499\\
$E^{miss}_T <$ 60 GeV & 4003\\ 
$\Delta \phi(b_i,b_j) >$ 1.5 & 1268\\
$\chi^{2} <$ 11 & 829\\
$\langle \Delta \phi(b,\Wboson) \rangle <$ 2 & 780\\ 
\bottomrule
\end{tabular}
\caption{The cut-flow table for all-hadronic sample with $m_t$ = 172.5 GeV. 
The initial number of events is determined using the ATLAS integrated luminosity $L = 20.2$ fb$^{-1}$ and the production cross section as calculated using  \texttt{MadGraph5\_aMC@NLO}.} 
\label{table:cutflowHad} 
\end{table}

\subsection{${R_{32}}$ Templates}
\label{sec:alljclose}
\begin{figure}[t]
\begin{center}
\includegraphics[width=0.98\linewidth]{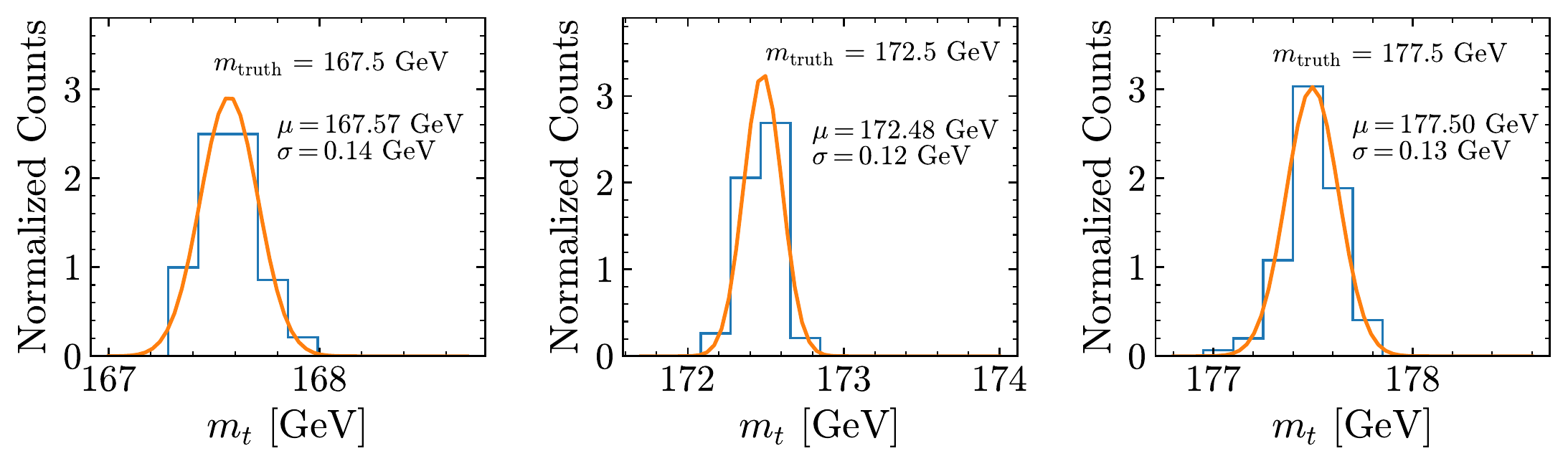}%
\caption{All-hadronic template closure tests for $m_t$=167.5 GeV, 172.5 GeV and 177.5 GeV.  
The distribution of bootstrapped results are given by the histogram, along with an accompanying Gaussian fit, whose mean and standard deviation are inset within each panel.
} 
\label{fig:alljclose}
\end{center}
\end{figure}

The observable $R_{32}$ is defined as the ratio of the three-jet mass to the di-jet mass, where the three-jet is a proxy for the top decay and di-jet is associated with the $W$ boson decay. 
One reason this observable is chosen for building the templates is that it partially reduces the systematic errors due to uncertainties in the JES. 
Following ATLAS, we fit the $R_{32}$ distributions for each of the 5 MC $m_t$ choices to the sum of a Landau function as defined in \cref{eqn:Landau}, and a Novosibirsk function defined as 
\begin{equation}
F(x)=N \exp\bigg(-\frac{1}{2\,\sigma_0^{2}} \,\log^{2}\bigg(1-\frac{x-x_p}{\sigma_E}\,\eta\bigg)-\frac{\sigma_0^{2}}{2}\bigg)\,,
\end{equation}
with
\begin{equation}
\sigma_0 = \frac{1}{\sqrt{\log 4}} \sinh^{-1}\big(\sqrt{\log 4}\, \eta\big)\,
\end{equation}
where $x_p$ is the peak, $\sigma_E$ is the width and $\eta$ is the asymmetry tail factor.
As described in \cref{sec:toy} above, we account for the statistical uncertainties associated with having a finite data set by bootstrapping 100 samples which are taken to be 3/5 of the full data.
We then repeat the $R_{32}$ fit for each of these datasets, and use these as input to generate a template as a function of $m_t$.
We then test that this procedure closes, which gives the histogram plotted in~\cref{fig:alljclose}.
Fitting these distributions to a Gaussian gives a quantitative measure of the closure goodness in the form of the mean and standard deviation given in each panel.

\begin{figure}[t]
\begin{center}
\includegraphics[width=0.85\linewidth]{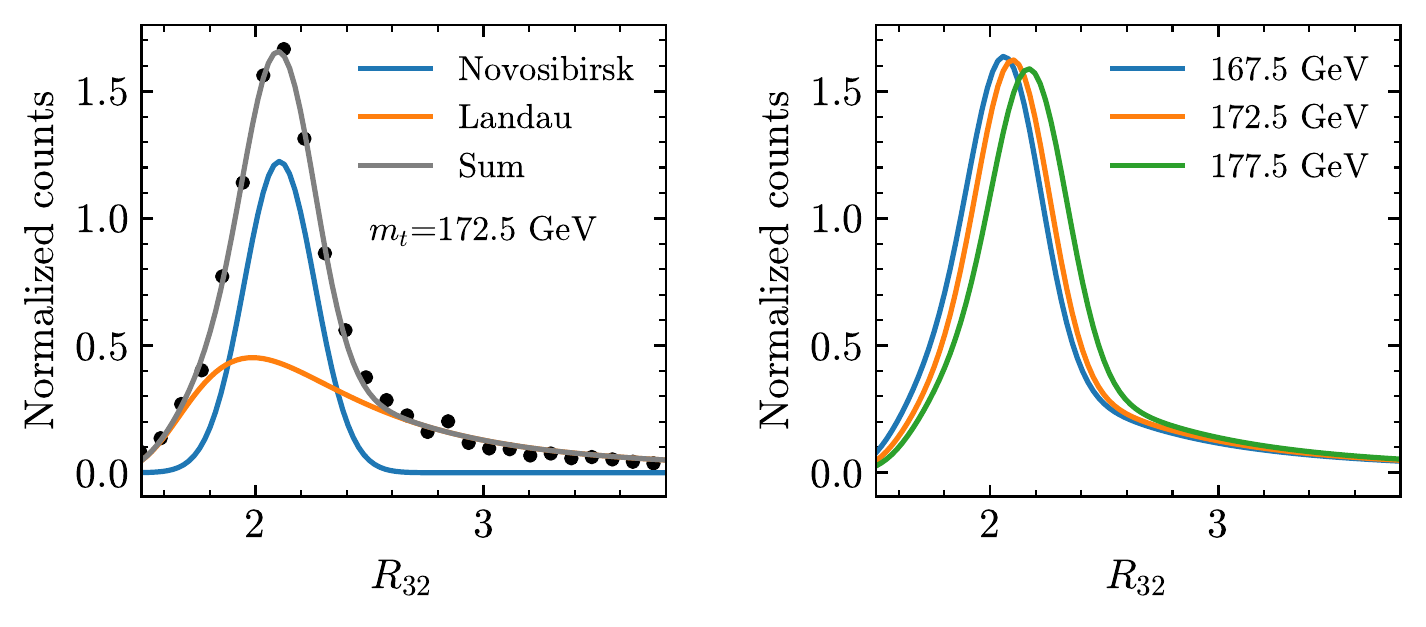}%
\caption{
The left panel gives the individual contributions of the Novosibirsk and Landau function using the best fit parameters for the $m_t$=172.5 GeV sample.
The right panel gives the $R_{32}$ template fits for $m_t$=167.5 GeV, 172.5 GeV, and 177.5 GeV, respectively, in the all-hadronic channel. 
} 
\label{fig:alljtemp}
\end{center}
\end{figure}

\section{The Di-leptonic Channel}
\label{sec:dilep}
In the di-leptonic channel, each of the $W$ bosons decays into a charged lepton and a neutrino. 
The final state is characterized by two $b$-jets, two leptons ($e$ or $\mu$), and \MET\!.  
One advantage of this channel is that the background is relatively low, especially in the $e \mu$ final state where there is no contribution from $Z$ boson decays.
Some drawbacks are that the branching ratio is only 10.5$\%$, and that it is not possible to reconstruct the top mass directly since there are two neutrinos that contribute to the \MET\!.
 \subsection{Pre-selection Cuts}
 \label{sec:cut2}
The physics object definitions are given as follows.
Electron candidates are required to have a transverse momentum of $p_T >$ 25 GeV and a rapidity $|\eta| < 2.47$ excluding range (1.37, 1.52). 
Muon candidates must satisfy $p_{T} >$25 GeV and $|\eta| < 2.5$. 
Muons must additionally satisfy an isolation requirement:  muons within a $\Delta R$ = 0.4 cone about the axis of a jet that has $p_T >$ 25 GeV are not considered.
Jets must satisfy an isolation requirement: events with jets that lie within a $\Delta R$ = 0.2 cone about the axis of an electron candidate are removed. 
Then, an electron isolation requirement discards events where electrons are found within $\Delta R$ = 0.4 cone about any of the remaining jets. 
The $b$-tagging efficiency is set to $0.7$, and rejection factors of $5$ and $137$ are taken for $c$ and light-flavor quarks respectively.

Now that we have defined our objects, we will walk through the pre-selection requirements.
Events are required to have a signal from the single-electron or single-muon trigger and at least one primary vertex with at least five associated tracks (we assume the trigger efficiency is 100\% for events that have an isolated electron or muon). 
An event must have exactly two oppositely charged leptons, where at least one of them must match the object that fired the corresponding trigger. 
In the same lepton flavor channels, $\MET > 60$ GeV is required. 
The invariant mass of the lepton pair is must be $m_{\ell \ell} >$ 15 GeV, excluding a window within 10 GeV of the $Z$ boson mass. 
In the different lepton flavor channels, the scalar sum of $p_T$ of the two selected leptons and all jets is required to be larger than 130 GeV.
There must be at least two valid jets, and at least one of these jets must be $b$-tagged. 
Finally, a cut on $p_{T_{\ell b}} > 120$ GeV is required.\footnote{
To compute $p_{T_{\ell b}}$, we need to pair up the $b$ with one of the leptons.
ATLAS does this using a multi-variate approach (MV): the two jets carrying the highest MV1 weight are taken as the two $b$-jets originating from the decays of the two top quarks. 
In our analysis, we simply obtain the $b$-jets by selecting the two $b$-tagged jets with the highest and second highest $p_T$. 
If there is only 1 $b$-tagged jet in an event, then we take the second $b$ to be the un-tagged jet with the highest $p_T$. 
Whichever way of pairing up the lepton and the $b$-jet, gives a lower invariant mass is used to calculate $p_{T_{\ell b}}$, and the observable used in the template procedure  $m_{\ell b}$.
} 

\begin{figure}[t]
\begin{center}
\includegraphics[width=0.85\linewidth]{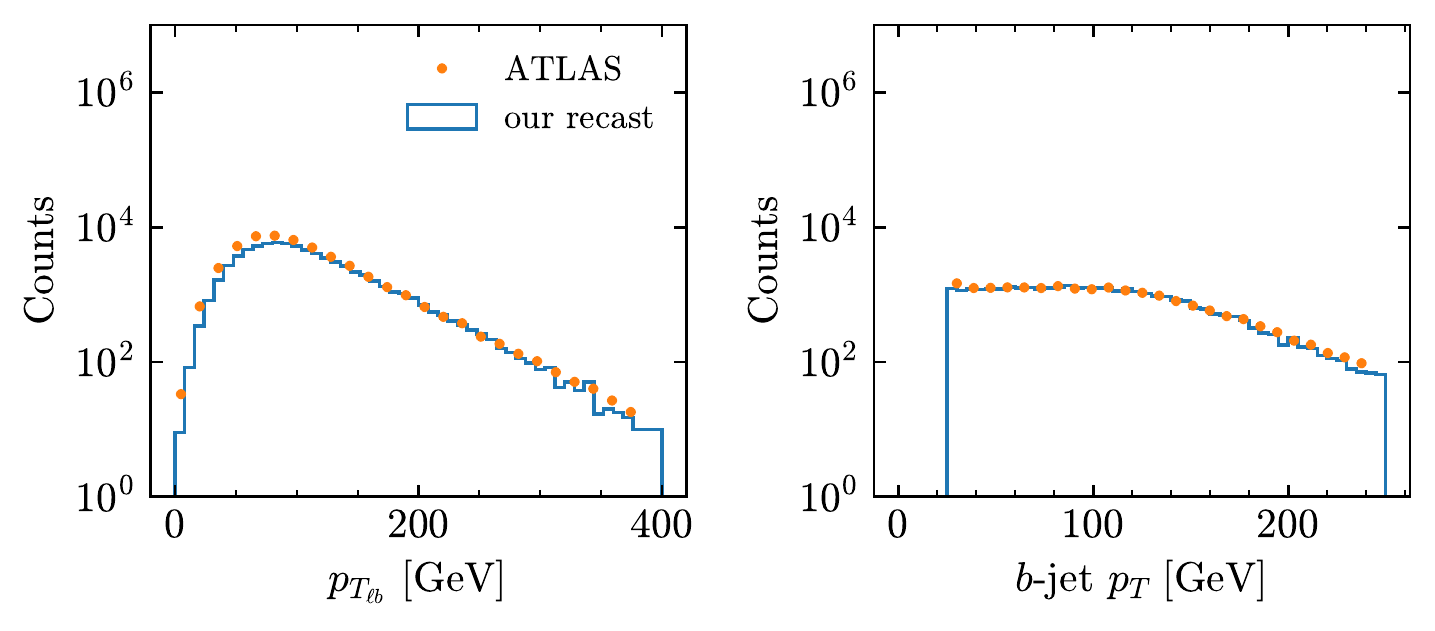}%

\caption{
Two validation plots are provided for the di-leptonic channel.  
The $p_{T_{\ell b}}$ distribution is given in the left panel, and the $b$-jet $p_T$ distribution is given on the right, assuming $m_t=172.5$ GeV. 
The blue distributions correspond to our simulation and the orange dots represent the ATLAS results~\cite{Aaboud:2016igd}, which have been rescaled to so that the normalizations agree. 
} 
\label{fig:leppre}
\end{center}
\end{figure}

\begin{table}[t]
\centering 
\renewcommand{\arraystretch}{1.3}
\setlength{\tabcolsep}{20 pt}
\begin{tabular}{l r r r} 
\toprule
Cut &\multicolumn{1}{c}{$e\,e$} & \multicolumn{1}{c}{$\mu\, \mu$} & \multicolumn{1}{c}{$e\,\mu$}\\[0.5ex] 
\hline 
Before cuts & 7796 & 13151& 20229\\ 
$\MET >$ 60 GeV & 3993 & 6784 & \multicolumn{1}{c}{$-$}\\
$m_{\ell \ell} >$ 15 GeV & 3942 & 6713 & \multicolumn{1}{c}{$-$}\\
$>$ 10 GeV from Z pole & 3365 & 5734 & \multicolumn{1}{c}{$-$}\\
$p_{T}(e+\mu+jets) >$ 130 GeV &  \multicolumn{1}{c}{$-$} & \multicolumn{1}{c}{$-$} & 19705\\ 
$\geq$ 2 valid jets & 2818 & 4783 & 16812\\
$\geq$ 1 b-tagged jet & 2450 & 4143 & 14628\\ 
\bottomrule 
\end{tabular}
\caption{
Cut-flow table for di-leptonic sample with $m_t$ = 172.5 GeV, separated by the flavor of leptons in the final state. 
The initial number of events is determined using the ATLAS integrated luminosity $L = 20.2$ fb$^{-1}$, and the production cross section is calculated using \texttt{MadGraph5\_aMC@NLO}.
} 
\label{table:cutflowlep} 
\end{table}

 \subsection{$m_{\ell b}$ Templates}
\label{sec:dilepclose}
\begin{figure}[t]
\begin{center}
\includegraphics[width=0.98\linewidth]{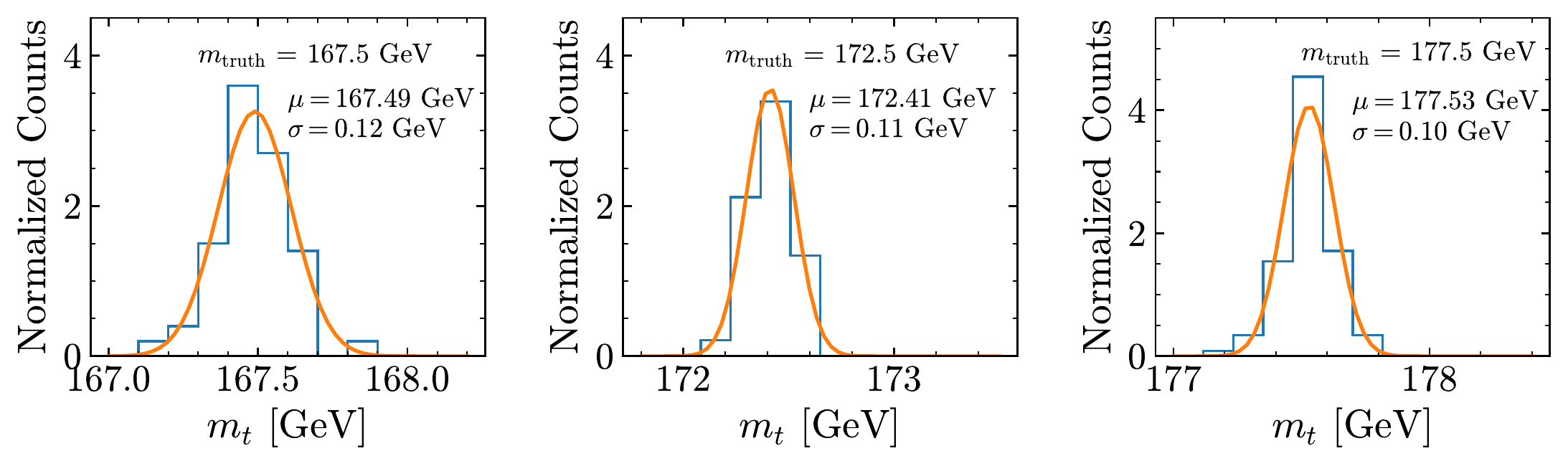}%
\caption{
Di-leptonic template closure tests for $m_t$=167.5 GeV, 172.5 GeV and 177.5 GeV.  
The distribution of bootstrapped results are given by the histogram, along with an accompanying Gaussian fit, whose mean and standard deviation are inset within each panel.
} 
\label{fig:lepclose}
\end{center}
\end{figure}

%
In \cref{fig:leppre}, we show two distributions computed using our samples after the pre-selection cuts have been applied, along with the comparisons to those given by ATLAS. 
The observable $m_{\ell b}$ is used to generate templates, where the parametric fit is now chosen to be a Landau function as defined in \cref{eqn:Landau} and a Gaussian.  
As described in \cref{sec:toy} above, we account for the statistical uncertainties associated with having a finite data set by bootstrapping 100 samples which are taken to be 3/5 of the full data.
We then repeat the $m_{\ell b}$ fit for each of these datasets, and use these as input to generate a template as a function of $m_t$.
We then test that this procedure closes by fitting the resulting histograms and comparing the fitted mean to the input $m_{\textrm{truth}}$ value, as shown in~\cref{fig:lepclose} .

\section{More on the Semi-leptonic Channel}
\label{sec:semi2}

This section provides extra validation information for the semi-leptonic channel. 
In Sec~\ref{sec:CutsSemiLep}, we summarized the pre-selection cuts for ATLAS's semi-leptonic analysis.
The detailed cutflow table is given by Table~\ref{table:cutflowsemi}. 

\begin{table}[t]
\centering 
\renewcommand{\arraystretch}{1.2}
\setlength{\tabcolsep}{20 pt}
\begin{tabular}{l r} 
\toprule
Cut & \multicolumn{1}{c}{$e$} \\ [0.5ex] 
\hline 
Before cuts & 89162  \\
1 lepton & 19437 \\ 
\MET $>$ 30 GeV & 14926 \\
$m_{T}^{W} >$ 30 GeV  & 13075 \\
$\geq4$  jets & 7793 \\
2 $b$-tagged jets & 3173 \\ 
\hline 
\end{tabular}
\hspace{20pt}
\begin{tabular}{l  r} 
\toprule
Cut &  \multicolumn{1}{c}{$\mu$}\\ [0.5ex] 
\hline 
Before cuts &  89162 \\
1 lepton & 25907\\ 
\MET $>$ 20 GeV &  23089\\
$\MET+m_{T}^{W} >$ 60 GeV &  21791\\
$\geq4$  jets &  12883\\
2 $b$-tagged jets &  5262\\ 
\hline 
\end{tabular}
\caption{Cutflow table for semi-leptonic sample with $m_t$ = 172.5 GeV, separated by the identity of the lepton. 
The initial number of events is determined using the ATLAS integrated luminosity $L = 20.2$ fb$^{-1}$, and the production cross section as calculated using \texttt{MadGraph5\_aMC@NLO}.
} 
\label{table:cutflowsemi} 
\end{table}

\begin{figure}[t]
\begin{center}
\includegraphics[width=0.98\linewidth]{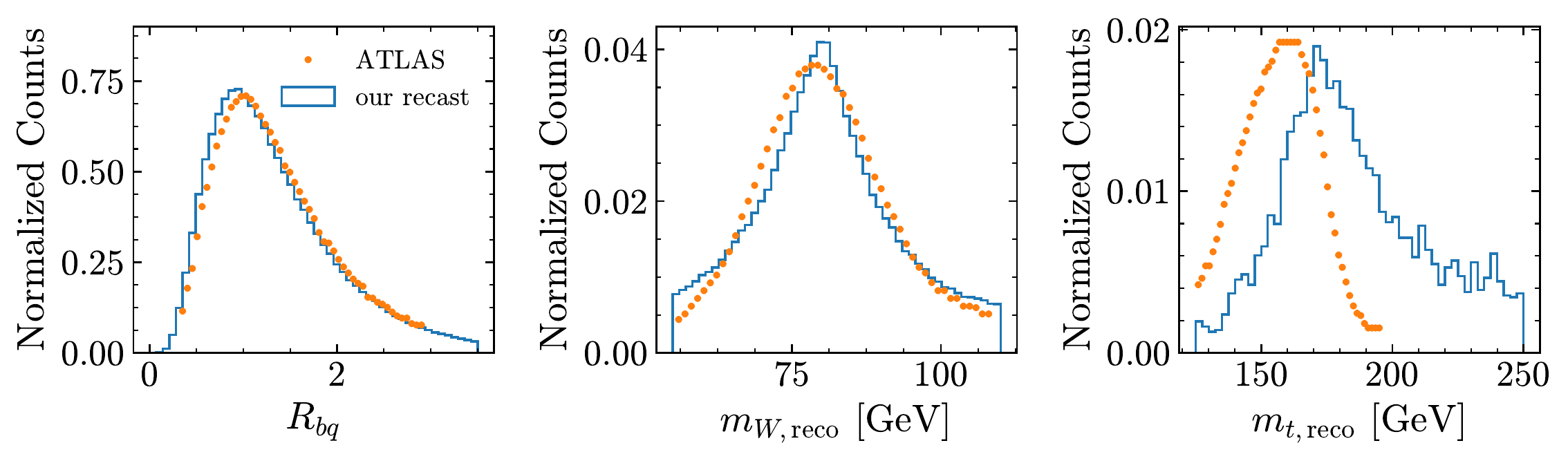}%
\caption{
This figure provides distributions of the three different observables used in the template analysis for the measurement of the top quark mass by ATLAS in the semi-leptonic channel. 
The panels from left to right show $R_{bq}$ which is defined in \cref{eq:Rbq}, $m_{W,\rm{reco}}$, and $m_{t,\rm{reco}}$, where the later two are the determined from maximizing the event-by-event likelihood defined in \cref{eqn:like}. 
The distributions for our analysis are shown by the solid blue histogram, while the orange dots are provided by ATLAS in~\cite{Aaboud:2018zbu}.  
We observe good agreement for the $m_{W,\rm{reco}}$ and $R_{bq}$ distributions, but we are unable to reproduce the \mtreco distribution. 
}
\label{fig:semimatch}
\end{center}
\end{figure}

Distributions of the observables used by ATLAS in this channel are shown in \cref{fig:semimatch}.
Note that we do not include the transfer functions in our likelihood, which specifically affects the \mtreco distribution, as explained in \cref{sec:mod}.

To make our proposed two-dimensional templates, we generate parton-level events with \texttt{MadGraph5\_aMC@NLO} that are subsequently passed to \texttt{Pythia8} for showering and hadronization, and then to \texttt{Delphes} to model detector effects.  
Five million events are generated for each of five choices for the top mass; a table providing the number of events that pass the preselection cuts is given in \cref{table:cutflowsemi}.
For each event, we use the likelihood defined in \cref{eqn:like} to chose the assignment of the jets, $b$-jets, $p_{z,\nu}$, and $m_{W,\rm{reco}}$. 
The resulting two-dimensional distributions in the $m_t^{\rm{had}}$ versus $m_t^{\rm{lep}}$  plane are then fit using the parametric function defined in \cref{eqn:2dtemplate}.

\begin{figure}[h]
\begin{center}
\includegraphics[width=0.9\linewidth]{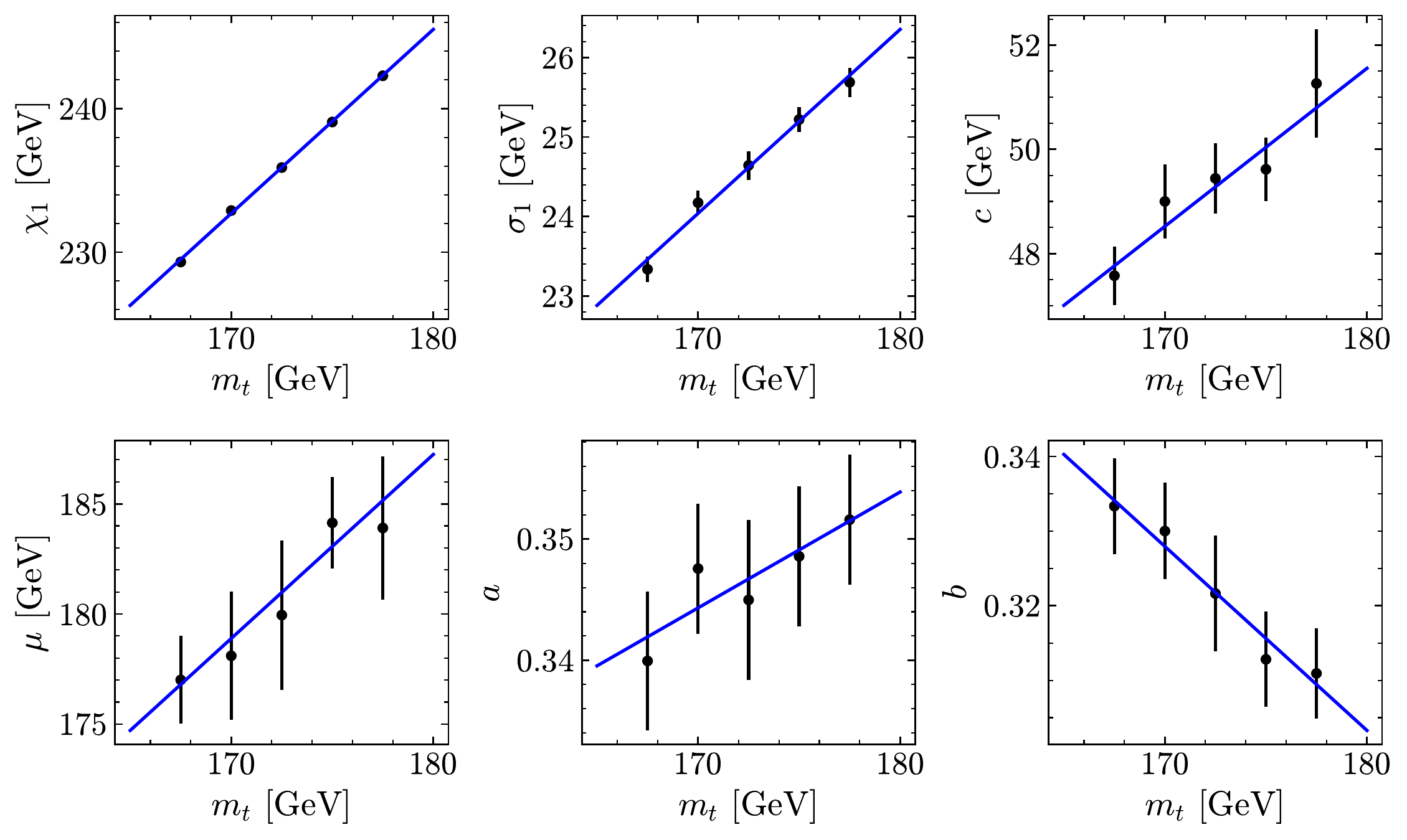}%
\hspace{30pt}
\includegraphics[width=0.65\linewidth]{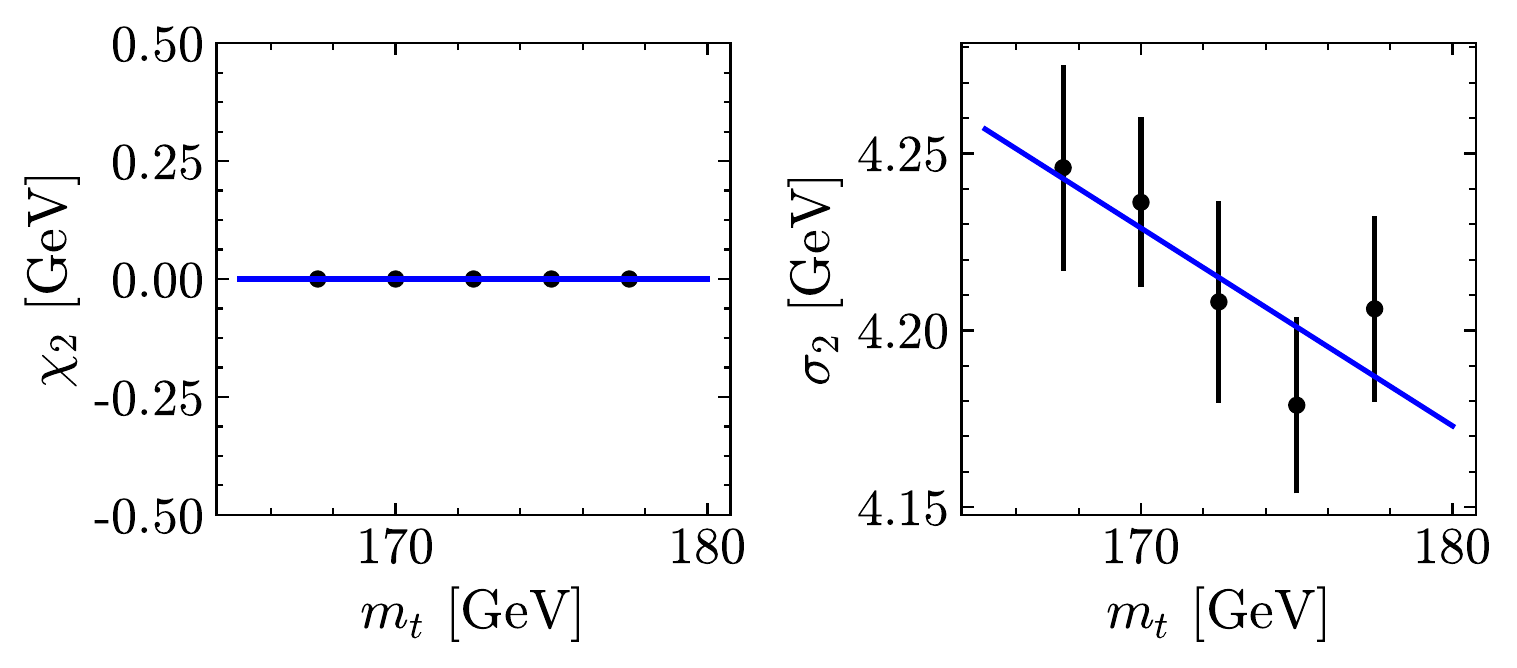}%
\caption{Here we provide the mean and standard deviation of the best fit parameters for the 100 samples derived using the bootstrap procedure.  
These distributions are then fit to a line as a function of $m_t$ in order to generate the semi-leptonic template.
} 
\label{fig:ljlinear}
\end{center}
\end{figure}

In order to follow the procedure discussed in \cref{sec:tpl}, we would like to have a set of $\sim 100$ statistically independent samples to work with.
However, it is computationally to expensive to re-generate the 5 million events many times. 
Therefore, we circumvent this issue using the statistical bootstrap, see \emph{e.g.}~\cite{wiki:bootstrap}.
Specifically, we random draw 3/5 of the 5 million events 100 times, allowing for replacement such that some events can be drawn more than once. 
We then find the best fit using our parametric function to each bootstrapped data set, providing us with an ensemble of 100 best fit parameters. 
The results are shown in \cref{fig:ljlinear}, where the points and error bars show the mean and standard deviation, respectively, of the best fit value for each parameter at each of the five top mass choices. 
Finally, we fit a line to each of these parameters as a function of the top mass, which is the input needed to define our two-dimensional template as a function of a single top mass parameter.

\begin{figure}[h]
\begin{center}
\includegraphics[width=0.98\linewidth]{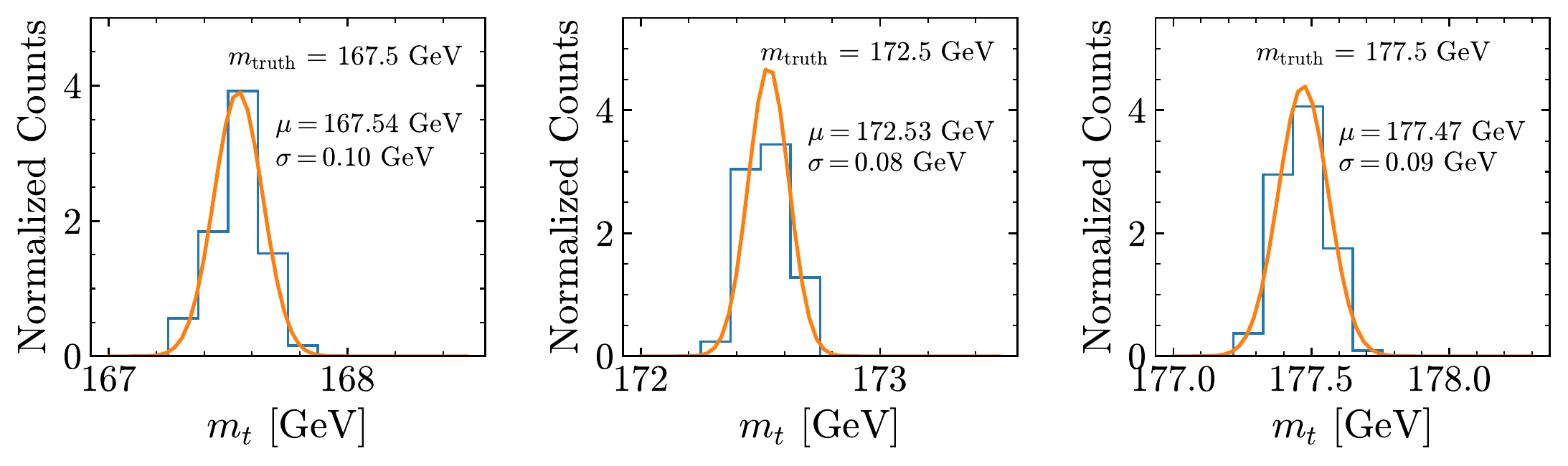}%
\caption{
The semi-leptonic template closure tests for $m_t$=167.5 GeV, 172.5 GeV and 177.5 GeV, using our modified approach.
The distribution of bootstrapped results are given by the histogram, along with an accompanying Gaussian fit, whose mean and standard deviation are inset within each panel. 
}
\label{fig:ljclose}
\end{center}
\end{figure}

The closure test is performed by taking another independent bootstrapped sample of 3 million events and fitting this data to our template to derive a best fit value of $m_t$.
This is repeated 100 times for each of the five truth top mass choices. 
The results are shown in \cref{fig:ljclose}, we show the results of this closure test which extracts the correct mass with a relatively small statistical uncertainty $\sim 0.1 \gev$.

\section{Combining Measurements}
\label{Sec:Combined}
%
Since we are interested in the global impact of stop contamination, it is useful to develop a simple framework for combining the measurements made in multiple independent channels.
In particular, the results in \cref{fig:SMCombination} of the main text show parameters which best fit the mass measurements combining the three channels. 
In the SM only assumption (left panel of \cref{fig:SMCombination}), the top mass which minimizes the $\chi^2$ error is $m_t^{\rm{comb}}=172.83 \GeV$, and the uncertainty band is determined by finding the contour where the $\chi^2$ is larger than the minimum by 1.0. 
We note that both the all-hadronic and the semi-leptonic central values lie outside this best-fit uncertainty band. 
The regions shown in the BSM parameter space are computed in a similar fashion, see the middle and right panels of \cref{fig:SMCombination}.

These regions only show the best-fit, and in particular they do not tell us how good the fit is. 
Therefore, it is amusing to ask if light stops can actually improve the fit to distributions measured by ATLAS.
To perform a quantitative test, we compute the likelihood ratio for observing the measured values in the three channels in the BSM scenarios as compared to the SM-only assumption. 
The test statistic is given by
\begin{align}
\lambda_{\rm{comb}} = -2\, \log \frac{\prod_{i\in{\rm{channels}}} P_{\rm{Gaussian}} \Big(m^{\rm{pred}}_{t,i};~m_{t,i}^{\rm{Obs}},\sigma^{\rm{Obs}}_i \Big)}
{\prod_{i\in\rm{channels}} P_{\rm{Gaussian}} \Big(m_t^{\rm{comb}};~m_{t,i}^{\rm{Obs}},\sigma^{\rm{Obs}}_i \Big)}\,,
\label{eqn:Lambda}
\end{align}
where $m_{t,i}^{\rm{Obs}}$ and $\sigma^{\rm{Obs}}_i$ are the channel specific measurement and uncertainty given by ATLAS, and $m^{\rm{pred}}_{t,i}$ is the value predicted in the BSM model for a given Monte Carlo \mt, \mstop, and \mnino.
This test statistic is constructed so that when $\lambda_{\rm{comb}} > 0$ the SM is a better fit, while when $\lambda_{\rm{comb}} < 0$ the BSM scenario is preferred.

The values of $\lambda_{\rm{comb}}$ computed in the $\mstop$ - $m_t$ plane are shown in \cref{Fig:BSM_to_SM}. 
The red regions correspond to values of $\lambda_{\rm{comb}} > 0$, indicating that the SM alone provides a better fit to the data. 
We gray out any parameter space with $\lambda_{\rm{comb}} > 1$ for brevity, since this region has a much stronger preference for the SM alone (and is of course additionally constrained by direct searches for stops). 
The white regions have a similar fit between the models, giving $\lambda_{\rm{comb}} = 0$. 
Intriguingly, we find a small region in the $\mnino=20\gev$ panel with $\lambda_{\rm{comb}} < 0$. 
However, our analysis yields that this parameter point is a mere 0.1~$\sigma$ more consistent with the data than the SM alone. 
We \emph{do not} take this to be evidence for a BSM contribution to the top mass measurements.

\clearpage

\addcontentsline{toc}{section}{\protect\numberline{}References}%
\bibliographystyle{utphys}
\bibliography{reference}

\end{document}